\documentclass[a4paper,fleqn,usenatbib]{mnras}

\usepackage{graphicx}   
\usepackage{amsmath}    
\usepackage{amssymb}    
\usepackage{subfig}

\def\dgc{\ensuremath{d_{GC}}}

\def\farcsec{\hbox{$.\!\!^{\prime\prime}$}} 
\def\rlim{\ensuremath{r_{\mathrm{lim}}}}
\def\muVmax{\ensuremath{\mu_{\mathrm{V,max}}}}
\def\muVlim{\ensuremath{\mu_{\mathrm{V,lim}}}}

\usepackage[T1]{fontenc}
\usepackage{ae,aecompl}

\title[dwarf galaxy remnants around 14 GCs]{A survey for dwarf galaxy 
remnants around fourteen globular clusters in the outer halo}
\author[Sollima et al.]{A. Sollima$^{1}$\thanks{E-mail:
antonio.sollima@oabo.inaf.it}, 
D. Mart\'{\i}nez Delgado$^{2}$, 
R. R. Mu\~{n}oz$^{3}$, 
J. A. Carballo-Bello$^{4,5}$,
\newauthor
D. Valls-Gabaud$^{6}$, 
E. K. Grebel$^{2}$, 
F. A. Santana$^{3}$, 
P. C\^{o}t\'{e}$^{7}$, 
S. G. Djorgovski$^{8}$\\
$^{1}$ INAF Osservatorio Astronomico di Bologna, via Ranzani 1, 40127 Bologna, 
Italy\\
$^{2}$ Astronomisches Rechen-Institut, Zentrum f\"{u}r Astronomie der Universit\"{a}t Heidelberg, M\"{o}nchhofstr. 12-14, 69120 Heidelberg, Germany\\
$^{3}$ Departamento de Astronom\'{\i}a, Universidad de Chile, Camino El Observatorio 1515, Las Condes, 7591245 Santiago, Chile\\
$^{4}$ Instituto de Astrof\'{\i}sica, Facultad de F\'{\i}sica, Pontificia Universidad
Cat\'{o}lica de Chile, Av. Vicu\~{n}a Mackenna 4860, Macul, 8970117 Santiago, Chile\\
$^{5}$ Millenium Institute of Astrophysics, Camino El Observatorio 1515, Las Condes, 7591245 Santiago, Chile. Santiago, Chile\\
$^{6}$ LERMA, CNRS UMR 8112, PSL, Observatoire de Paris, 61 Avenue de
l'Observatoire, 75014 Paris, France\\
$^{7}$ National Research Council of Canada,
Herzberg Astronomy and Astrophysics Program, 5071 W. Saanich Road, Victoria V9E 2E7, BC, Canada\\
$^{8}$ Astronomy Department, California Institute of Technology, 1200 E. California Blvd, Pasadena, CA 91125,
USA\\
}

\date{Accepted XXX. Received YYY; in original form ZZZ}

\pubyear{2017}

\begin{document}
\label{firstpage}
\pagerange{\pageref{firstpage}--\pageref{lastpage}}
\maketitle


\begin{abstract}
We report the results of a systematic photometric survey of the peripheral regions
of a sample of fourteen globular clusters in the outer halo of the Milky
Way at distances $d_{GC}>25$ kpc from the Galactic centre. The survey is aimed at searching for the remnants of
the host satellite galaxies where these clusters could originally have been formed before being
accreted onto the Galactic halo. The limiting surface brightness varies
within our sample, but reaches $\muVlim=30-32$~mag arcsec$^{-2}$. 
For only two globular clusters (NGC 7492 and Whiting 1; already suggested
to be associated with the Sagittarius galaxy) we detect extended stellar 
populations that cannot be associated with either the clusters themselves or with 
the surrounding Galactic field population.
We show that the lack of substructures around globular clusters at these  
Galactocentric distances is still compatible with the predictions of cosmological
simulations whereby in the outer 
halo the Galactic globular cluster
system is built up through hierarchical accretion at early
epochs. 
\end{abstract}


\section{Introduction}
\label{intro_sec}

According to the most widely accepted cosmological model,
the {\em concordance} $\Lambda$ -- cold dark matter model 
\citep[$\Lambda$CDM][]{2016A&A...594A...1P}, galaxies like the Milky Way are mainly
formed by the hierarchical assembly of smaller subsystems \citep[][and references therein]{2015ApJ...799..184P,2016MNRAS.458.2371R}. 

A typical late accretion event would
involve a dwarf satellite that is progressively disrupted by the tidal pull
exerted by the parent galaxy. The stripped material
(gas/ stars / clusters / dark matter) is placed on orbits similar to that of the original
satellite, hence forming tidal tails that may form multiple filamentary wraps around the parent galaxy \citep[see][]{2005ApJ...619..807L}.
 Accretion events
in the inner halo at high redshift completed tens of orbits distributing their
remnants homogeneously in space, while satellites accreted more
recently and at large Galactocentric
distances are expected to leave compact remnants. 
Gas-rich mergers constitute also
the reservoir from which a homogeneous component of the halo could form \textsl{in situ} in the
densest central region of the Milky Way. 
Models of galaxy formation predict that
the haloes of Milky Way-like galaxies should therefore be characterised by a 
superposition of  stellar populations from these substructures with a degree of
discreteness that increases towards large Galactocentric distances  
\citep{2005ApJ...635..931B,2010MNRAS.406..744C}. 

The tidal features left by the hierarchical formation
process should be still observable in the outer halo and their detection 
provides an observational test of the $\Lambda$CDM
paradigm. For this reason, in recent years  huge observational 
efforts have been carried out 
 to search for overdensities in large-scale photometric surveys like the Sloan Digital Sky
Survey \citep[SDSS,][]{2009ApJS..182..543A}, the Two-Micron All Sky Survey 
\citep[2MASS,][]{ 2006AJ....131.1163S} and Pan-STARRS \citep{2016arXiv161205560C}.
 Truly spectacular examples  have been observed
 both within the Milky Way, \textsl{i.e.} the case of the Sagittarius dwarf galaxy
and its giant tidal stream 
\citep{1994Natur.370..194I,2003ApJ...599.1082M} and 
the "field of streams" 
\citep{2006ApJ...642L.137B}, 
in M31 \citep{2001Natur.412...49I,2009Natur.461...66M} 
 and also in nearby galaxies 
\citep[see][]{2010AJ....140..962M,2015MNRAS.446..120D}.
 Studies devoted to the
quantification of the amount of substructures in the halo have been performed 
by \cite{2008ApJ...680..295B}, \cite{2009ApJ...698..567S}, \cite{2011ApJ...738...79X} 
and \cite{2016ApJ...816...80J},
who indeed find that the halo is highly structured with an increasing clumpiness at large Galactocentric distances.

The outskirts of globular clusters (GCs) are
among the best places to look for remnants of the ancient satellites
predicted by the $\Lambda$CDM theory.
Indeed, in the classical \cite{1978ApJ...225..357S} 
 scenario of the formation of
the Galaxy, at least part of the outer halo GCs (at Galactocentric distances $d_{GC}>8$ kpc) 
formed in dwarf galaxies that were later accreted by the Milky Way.
This hypothesis is based on the absence in the group of GCs populating the 
outermost Galactic halo of the clear metal abundance gradient 
observed in the inner Galaxy. An updated view of this evidence comes 
from the distribution of GCs in the age-metallicity plane 
\citep{2009ApJ...694.1498M,2010MNRAS.404.1203F,2013MNRAS.436..122L}:  GCs located at distances $d_{GC}>15$ kpc tend to populate a
separate branch of this diagram, away from the locus of nearby GCs. 
Moreover, these clusters exhibit 
peculiar kinematical properties
(large, energetic orbits of high eccentricity), larger core radii, a higher
specific frequency of RR Lyr{\ae} stars 
\citep{2004MNRAS.355..504M}
 and most of them 
are distributed along an inclined disc that encompasses the Galaxy, 
together with the dwarf spheroidals and the ultra-faint dwarfs 
\citep{2012MNRAS.423.1109P,2012ApJ...744...57K}.
The extra-Galactic origin of these clusters is also
invoked to explain the bimodal metallicity distribution of GCs observed in
both spiral and elliptical galaxies 
\citep{1998ApJ...501..554C,2000ApJ...533..869C,2005ApJ...623..650K,2017MNRAS.465.3622R}.
 However, these trends do not provide a clear-cut proof of the origin
of these GCs. On the other hand, if these clusters did form within larger stellar systems,
 they are expected to be surrounded by stellar populations of their host
 galaxy, over a wide field.
Remarkably, recent studies have revealed the presence of streams in the 
surroundings of a few GCs in the inner halo 
\citep{2002ApJ...573L..19M,2003A&A...405..577B,2018MNRAS.474..683C}.
In particular, the Sagittarius galaxy, the most prominent accretion event 
in the Milky Way, is expected to contribute to the Galactic GC system with 
$5\div9$ clusters, lying today along its orbital path 
\citep{2010ApJ...718.1128L}. Among them, the presence of the Sagittarius stellar population has been found in the 
surroundings of NGC 7492 and Whiting 1 \citep{2014MNRAS.445.2971C}, at Galactocentric distances $d_{GC}>25$ kpc.
 The same scenario seems to be confirmed in M31 where all GCs beyond 30 kpc 
from the centre appear to be associated with streams 
\citep{2010ApJ...717L..11M, 2014MNRAS.442.2929V}.

In this paper, we use deep wide-field photometric observations of a sample of 14
Galactic GCs populating the distant halo of the Milky Way (at distances $\dgc>25$
kpc) to investigate the possible presence of stellar populations
arising from their hypothetical progenitor hosts.  
In Section~\ref{sec:obs} we describe the observational
material and the adopted data reduction techniques. 
In Section~\ref{sec:methods} we describe the 
methodology used to detect overdensities and the results
of its
application to our sample. Section~\ref{sec:models} is devoted to the comparison of our results 
with a set of cosmological simulations. We discuss our
results in Section~\ref{sec:discussion}.

\section{Observational material}
\label{sec:obs}

\begin{table}
 \centering
 \label{tab:table1}
  \caption{Summary of observations.}
  \begin{tabular}{@{}lccccr@{}}
  \hline
 Name & Telescope & Filter & $N_{exp}$ & $t_{exp}$ & FoV\\
      &           &        &           &    [s]      & [deg$^2$]\\ 
 \hline
 Pal 13    &    CFHT   & g &    6      &   360     &  0.98\\
           &           & r &    6      &   360     &  0.98\\
 NGC 7492  &    CFHT   & g &    6      &   120     &  0.94\\
           &           & r &    6      &   120     &  0.94\\
           &    Clay   & g &    5      &    90     &  0.21\\
           &           & r &    5      &   180     &  0.21\\
 Whiting 1 &    CFHT   & g &    6      &   300     &  0.95\\
           &           & r &    6      &   300     &  0.95\\
 NGC 6229  &    CFHT   & g &    6      &    90     &  0.91\\
           &           & r &    6      &    90     &  0.91\\
 AM 4      &    Clay   & g &    5      &    90     &  0.21\\
           &           & r &    5      &   180     &  0.21\\
Koposov 2  &    CFHT   & g &    6      &   500     &  0.94\\
           &           & r &    6      &   500     &  0.94\\
 NGC 5694  &    CFHT   & g &    6      &    60     &  0.95\\
           &           & r &    6      &    60     &  0.95\\
 NGC 7006  &    CFHT   & g &    6      &   240     &  0.95\\
           &           & r &    6      &   240     &  0.95\\
 Pal 14    &    CFHT   & g &    6      &   680     &  0.95\\
           &           & r &    9      &   680     &  0.95\\
 NGC 2419  &    CFHT   & g &    6      &   450     &  1.02\\
           &           & r &    6      &   450     &  1.02\\
 Eridanus  &    CFHT   & g &    6      &   270     &  0.94\\
           &           & r &    6      &   270     &  0.94\\
 Pal 3     &    CFHT   & g &    6      &   270     &  0.98\\
           &           & r &    6      &   270     &  0.98\\
           &    Clay   & g &    5      &    90     &  0.21\\
           &           & r &    5      &   180     &  0.21\\
 Pal 4     &    CFHT   & g &    6      &   440     &  0.96\\
           &           & r &    6      &   440     &  0.96\\
 AM 1      &    Clay   & g &    5      &    90     &  0.21\\
           &           & r &    5      &   180     &  0.21\\
\hline
\end{tabular}
\end{table}
 
\begin{figure*}
 \includegraphics[width=\textwidth]{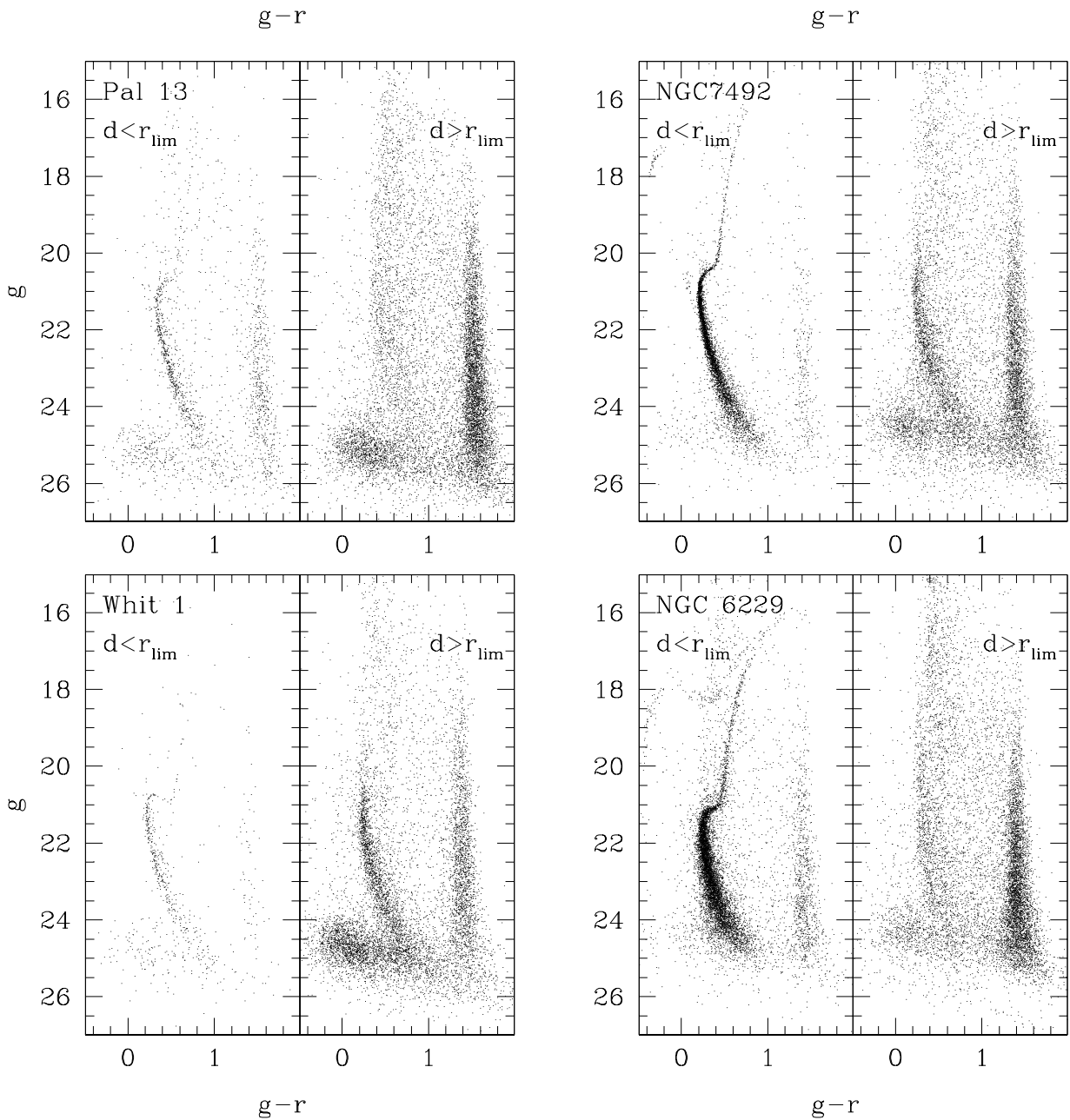}
 \caption{Colour-magnitude diagrams in  ($g,g-r$) of Pal 13, NGC 7492, Whiting 1, and NGC 6229. The left panels
 refer to the region within the estimated cluster limiting radius $\rlim$ (see
 \S~\ref{sec:methods}). The right panels
 refers to the region beyond $\rlim$. Only stars within the adopted sharpness
 threshold are shown.}
\label{fig:cmd}
\end{figure*}
 
\begin{figure*}
 \ContinuedFloat
 \includegraphics[width=\textwidth]{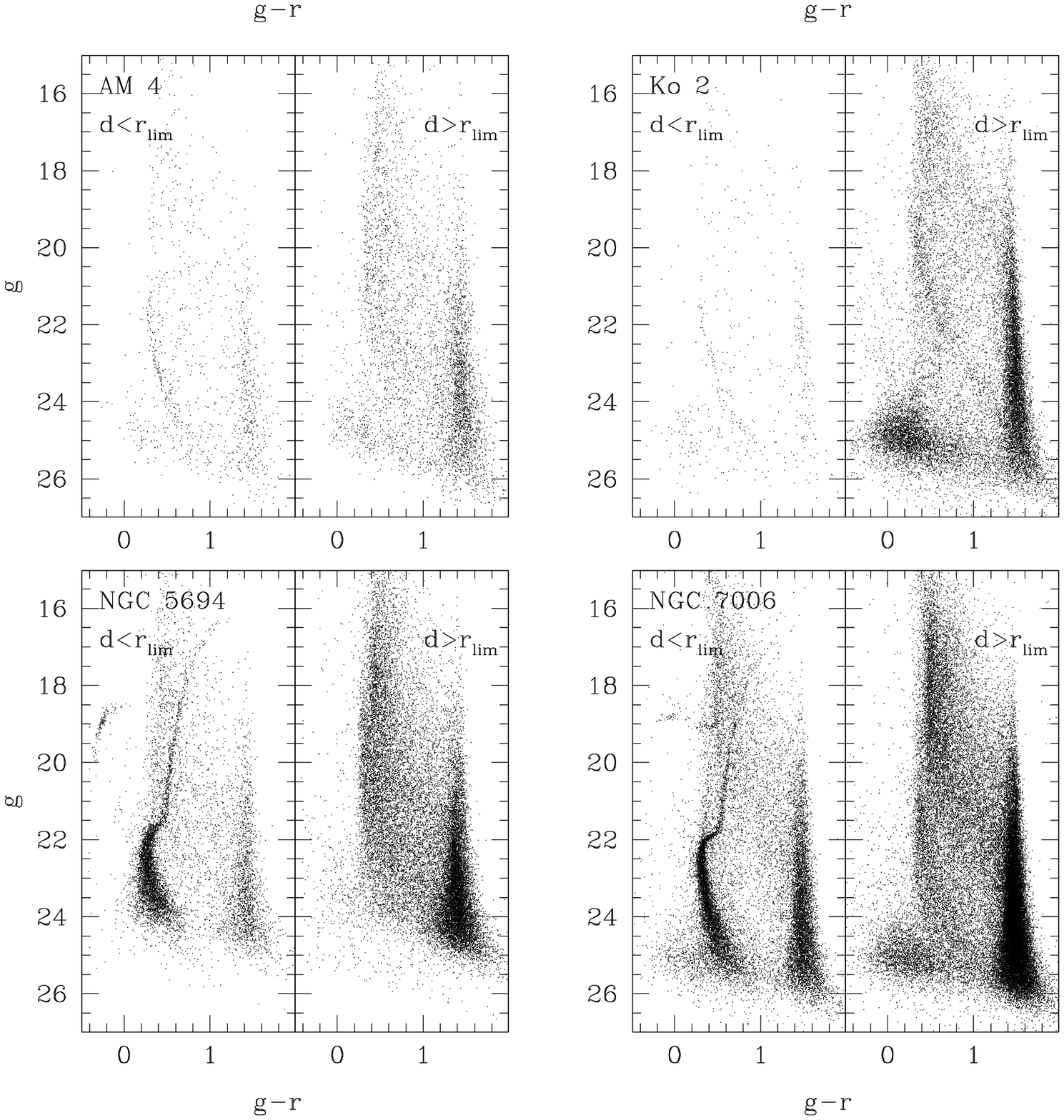}
 \caption{(Continued) Colour-magnitude diagrams for AM 4, Ko 2, NGC 5694, and NGC 7006.}
\end{figure*}
 
\begin{figure*}
 \ContinuedFloat
 \includegraphics[width=\textwidth]{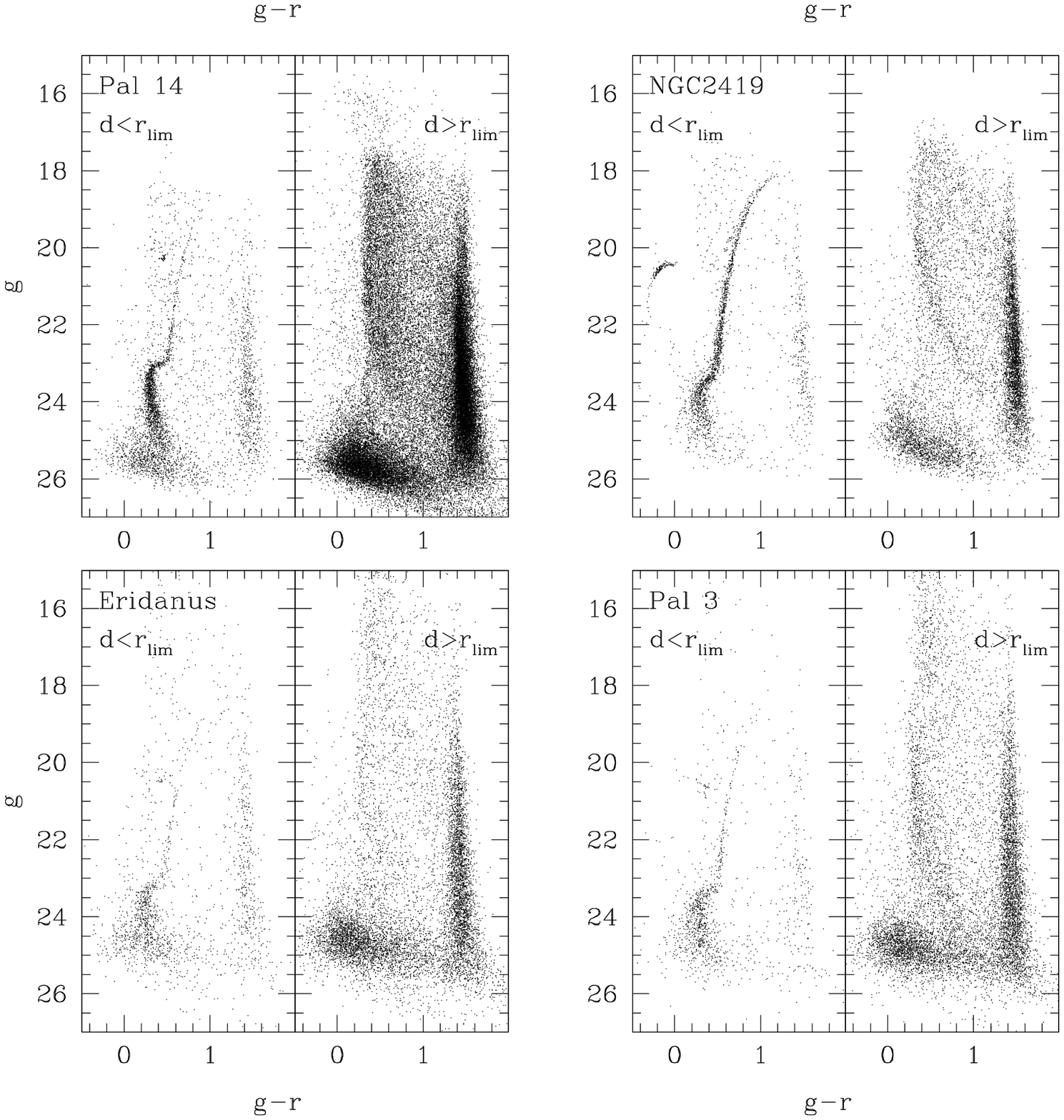}
 \caption{(Continued) Colour-magnitude diagrams for Pal 14, NGC 2419, Eridanus, and Pal 3.}
\end{figure*}
 
\begin{figure*}
 \ContinuedFloat
 \includegraphics[width=\textwidth]{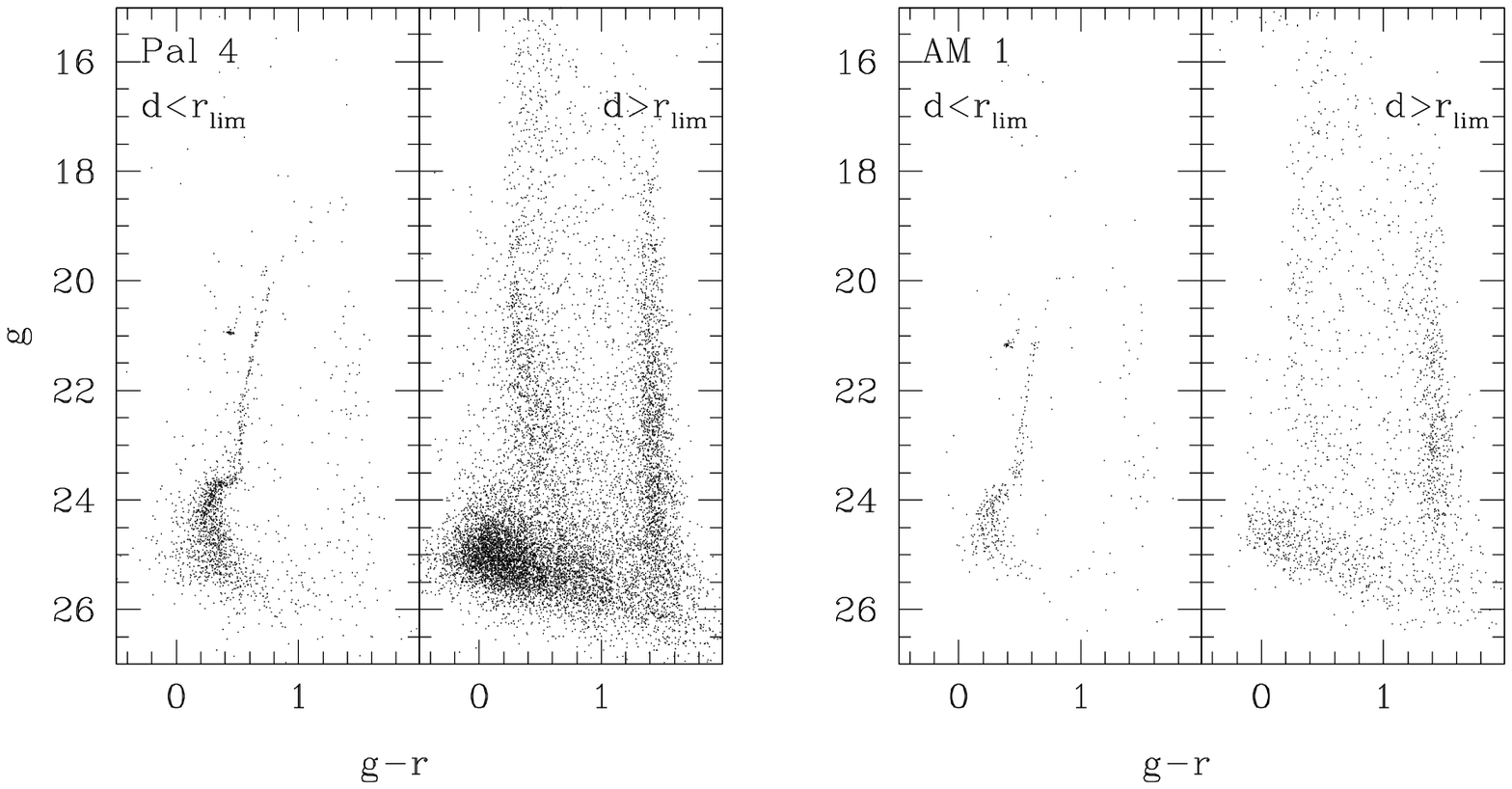}
 \caption{(Continued) Colour-magnitude diagrams for Pal 4 and
 AM 1.}
\end{figure*}

The GCs of our database were selected among those populating the outer halo of
the Milky Way at Galactocentric distances $\dgc>25$ kpc.
Among the 19 GCs listed in the Harris catalogue 
\citep[][2010 edition]{1996AJ....112.1487H}
 in
this distance range, we excluded two GCs at low Galactic latitude
$|b|<10^{\circ}$ (Pyxis and Pal~2) because of the severe effect of differential 
reddening affecting their colour-magnitude diagrams (CMDs), two GCs (NGC 5824 
and Pal 15) because their tidal extent exceeds the field of view of our data 
\citep[see also][]{2018MNRAS.473.2881K,2017ApJ...840L..25M} and one GC (Ko 1) 
because of its sparse number of stars, which did not allow us to perform our analysis. 
 The final sample hence consists of 14 GCs located in both hemispheres covering a
range of distances of 25$<\dgc/kpc<$125 from the Galactic centre.

The photometric dataset consists of a set of images collected with the imagers
Megacam (mounted at the CFHT, Canada-France Hawaii Telescope in Mauna Kea, 
Hawaii) and Megacam (at the Magellan II-Clay Telescope in Las Campanas, Chile)
as a part of a project
aimed at the characterisation of Galactic globular clusters 
\citep[][submitted]{2018Munoz}. 
This survey presents unprecedented photometric data quality for describing
an almost complete sample of outer halo satellites, which was already used in many published 
works \citep[see \textsl{e.g.}][]{2011ApJ...730L...6S,2011ApJ...726...47S,2011ApJ...743..167B,2013ApJ...774..106S,2015ApJ...805...51C}. The camera Megacam at CFHT is a prime focus mosaic 
camera consisting of 36 $2048\times4612$ pixel CCDs (a total of 340 megapixels), 
covering a $0^{\circ}.96 \times 0^{\circ}.94$ 
square degree field of view with a pixel-scale of 0\farcsec187 per pixels \citep{2003SPIE.4841...72B}.
Observations were performed in queue mode during 
semesters 2009-A, 2009-B, and 2010-A mostly in photometric conditions.
The average seeing was $\sim0\arcsec.8$. 
Clusters in the southern hemisphere were 
observed using the Megacam camera at Magellan, 
which is composed of 36 closely
packed CCDs assembled in an $18 432 \times 18 432$ pixels array. The
CCDs have 13.5 $\mu m$ square pixels that subtend
$0\farcsec08$ at the $f/5$ focus, yielding a $25\arcmin\times25\arcmin$ field of view. 
The average seeing of these observations was $\sim0\arcsec.9$.
For each cluster, multiple images were observed through the Sloan g$'$ and r$'$
filters, placing the cluster close to the centre of the camera field of view.
The log of the observations, including exposure times, number of exposures and
overall field of view for
each target cluster is listed in Table 1.
Standard photometric reduction (bias subtraction, flat-field, etc.) was carried using the IRAF\footnote{\texttt{IRAF} is distributed by the National 
Optical Astronomy Observatories, which are operated by the Association of 
Universities for Research in Astronomy, Inc., under cooperative agreement with 
the National Science Foundation.} task \texttt{ccdproc}.

The photometric analysis was carried out using the PSF-fitting algorithm of
the \texttt{Daophot/ALLFRAME} package \citep{1994PASP..106..250S}.
The brightest and most isolated stars were used to construct a model PSF
and to link PSF and aperture magnitudes.
Images were aligned and stacked to construct a high signal-to-noise image where
the source detection was performed. The photometric analysis 
was then done on the individual images and the resulting magnitudes were 
averaged. 
For the objects in common with the SDSS-DR7 catalogue \citep{2009ApJS..182..543A}, the
instrumental magnitudes were calibrated using the stars in common,
selecting the ones with $18<r_{\rm{sdss}}<21.5$ and $18<g_{\rm{sdss}}<22$ 
to avoid both saturated and large photometric uncertainty stars.
The objects that lay outside the SDSS footprint were calibrated using the average 
extinction terms, zero-points and the color-terms determined for the
GCs overlapping the SDSS imaging area and that were observed with the same instrument under the same observing 
conditions.
To limit the contamination from background galaxies, whose location in the CMD
partly overlaps the locus of the MS of a halo stellar population, we removed 
all stars with a sharpness parameter $|S|>0.3$. This value was chosen 
to include the bulk
of bona-fide cluster MS stars selected within $\Delta (g-r)<0.1$ about the MS 
ridge line and located within $R<\rlim$. This criterion has been found to provide 
the best compromise between completeness and contamination with a better 
discrimination power than other considered alternatives (e.g. PSF-fitting vs. 
aperture magnitude, etc.). 

The $(g-r)-g$ colour-magnitude diagrams (CMDs) of the 14 GCs of our sample are
shown in Fig.~\ref{fig:cmd}. They sample the cluster population from the tip of the red
giant branch to some $1 - 4$ mag below the main sequence (MS) turn-off depending on the
cluster distance, reaching a typical magnitude limit of $g\sim25.5$.
At large distances from the centres of the clusters the stellar populations of the
Galactic field dominate, showing a fairly homogeneous distribution of stars in
the blue portion of the CMD (at $g-r<1$; composed of the superposition of
disc/halo upper MS stars at different heliocentric distances) and a red plume
(at $g-r>1$; composed of low-mass MS dwarfs). 
In a few clusters (\textsl{e.g.} NGC 7492, Whiting 1), an overdensity of stars in the blue region of the CMD
resembling a MS feature is apparent.
At faint magnitudes (at $g>24$ and $g-r<0.6$) a large number of
background galaxies contaminate the CMD.
Artificial star experiments were performed on the 
science images of NGC 2419 and NGC 7006 
indicating a completeness level $>80\%$ at g$\sim$24 in both clusters, with a 
sharp decrease at fainter magnitudes.

\section{method}
\label{sec:methods}

\begin{table}
 \centering
 \label{tab:table2}
  \caption{Heliocentric and Galactocentric distance, Galactic latitude, estimated limiting radius,
  background surface brightness and index $Q$ of the clusters of our sample.}
  \begin{tabular}{@{}lcccccr@{}}
  \hline
 Name & $d_{\odot}$ & $d_{GC}$ & $b$ & $\rlim$ & $\muVlim$ & Q\\
      & [kpc] & [kpc] & [$^{\circ}$] & [$\arcmin$] & [mag arcsec$^{-2}$] &\\
 \hline
 Pal 13    &  26.0 & 26.9 & -42.70 & 11.2 & 30.96 & -7.93 \\
 NGC 7492  &  26.3 & 25.3 & -63.48 &  9.6 & 29.81 & 10.78 \\
 Whiting 1 &  30.1 & 34.5 & -60.76 &  5.8 & 30.29 & 17.41 \\
 NGC 6229  &  30.5 & 29.8 &  40.31 & 13.4 & 30.52 & -7.20 \\
 AM 4      &  32.2 & 27.8 &  33.51 &  5.9 & 30.06 & -4.92 \\
 Ko 2      &  34.7 & 41.9 &  25.54 &  4.8 & 31.27 & -11.50 \\
 NGC 5694  &  35.0 & 29.4 &  30.36 & 12.3 & 30.24 & -9.02 \\
 NGC 7006  &  41.2 & 38.5 & -19.41 & 14.9 & 29.89 & 0.75 \\
 Pal 14    &  76.5 & 71.6 &  42.19 &  9.6 & 30.69 &  \\
 NGC 2419  &  82.6 & 89.9 &  25.24 &  8.8 & 29.49 &  \\
 Eridanus  &  90.1 & 95.0 & -41.33 & 10.5 & 30.84 &  \\
 Pal 3     &  92.5 & 95.7 &  41.86 &  7.2 & 30.74 &  \\
 Pal 4     & 108.7 & 111.2 & 71.80 &  6.7 & 31.25 &  \\
 AM 1      & 123.3 & 124.6 & -48.47 &  3.6 & 31.63 &  \\
\hline
\end{tabular}
\end{table}

\begin{figure*}
 \includegraphics[width=\textwidth]{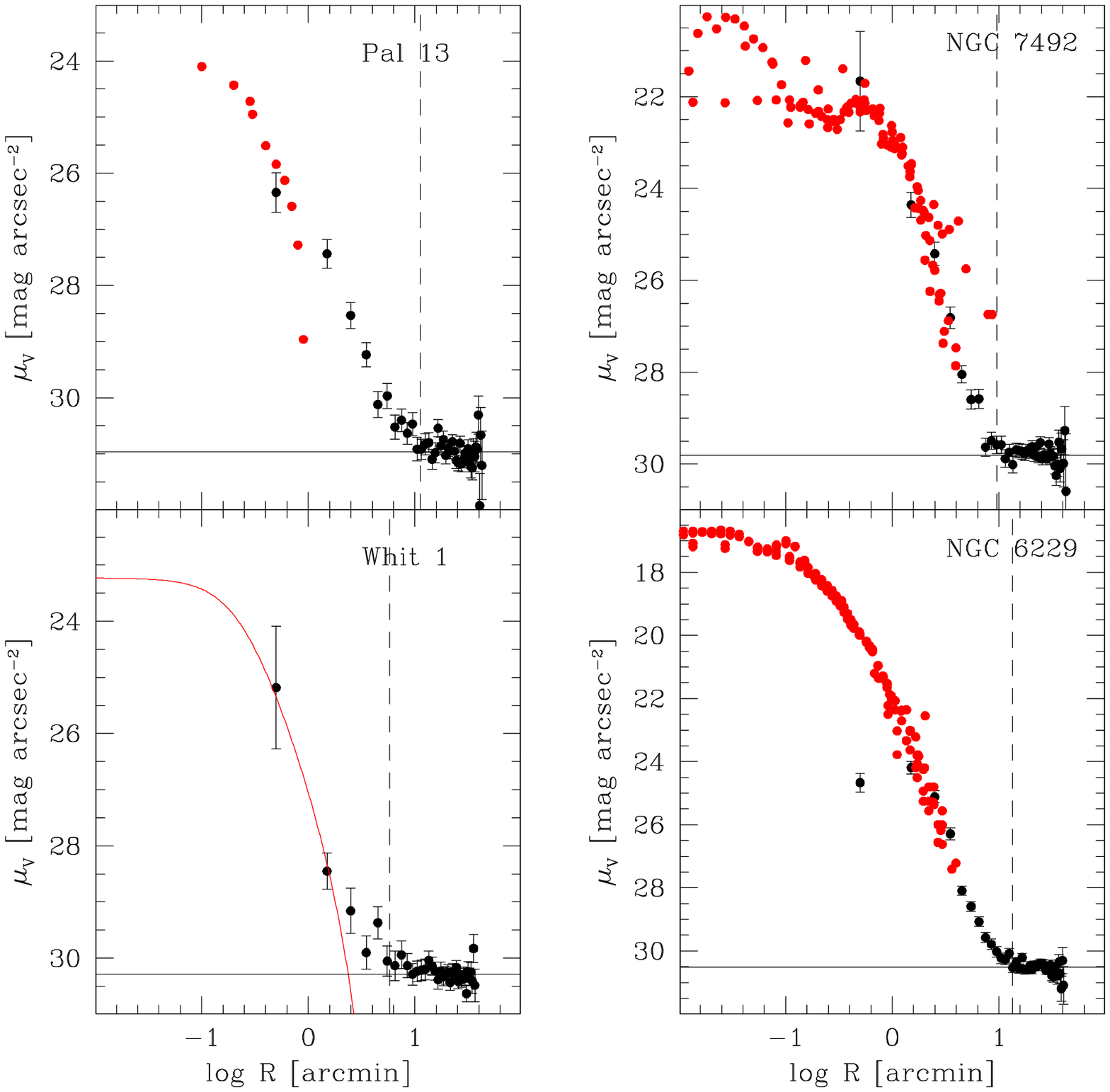}
 \caption{Surface brightness profiles of Pal 13, NGC 7492, Whiting 1, and NGC 6229. 
 Black points are measures from this work, red points (grey in the printed version of the paper)
 are from Trager et al. (1995). The solid red curve represents the best-fitting King (1966) 
  model. The adopted 
 limiting radius $\rlim$ and the background surface brightness 
 $\muVlim$
 are marked by the vertical dashed line and the horizontal solid line, respectively.}
\label{fig:profile}
\end{figure*}
 
\begin{figure*}
 \ContinuedFloat
 \includegraphics[width=\textwidth]{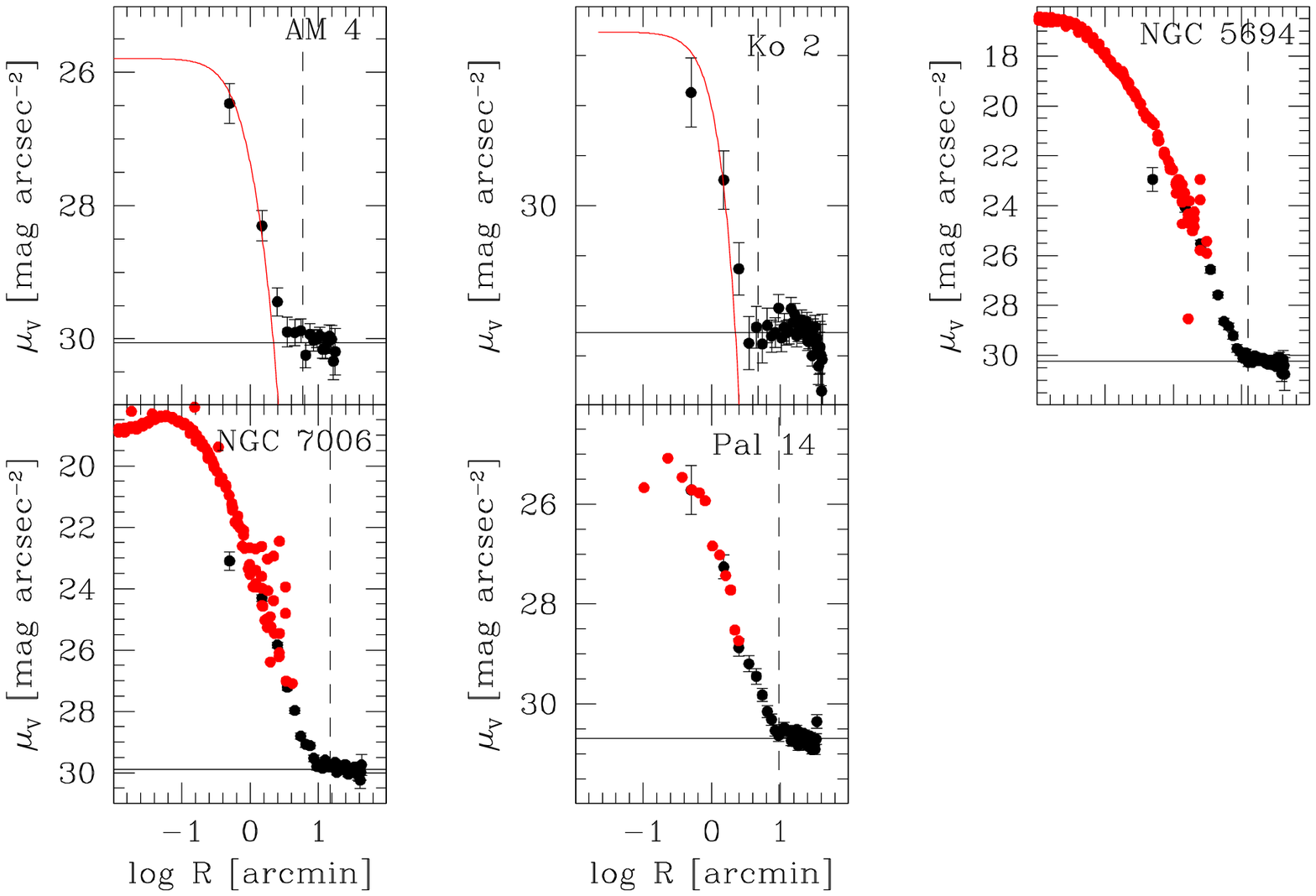}
 \caption{(Continued) Surface brightness profiles of AM 4, Ko 2, NGC 5694, NGC 7006 and Pal 14.}
\end{figure*}
 
\begin{figure*}
 \ContinuedFloat
 \includegraphics[width=\textwidth]{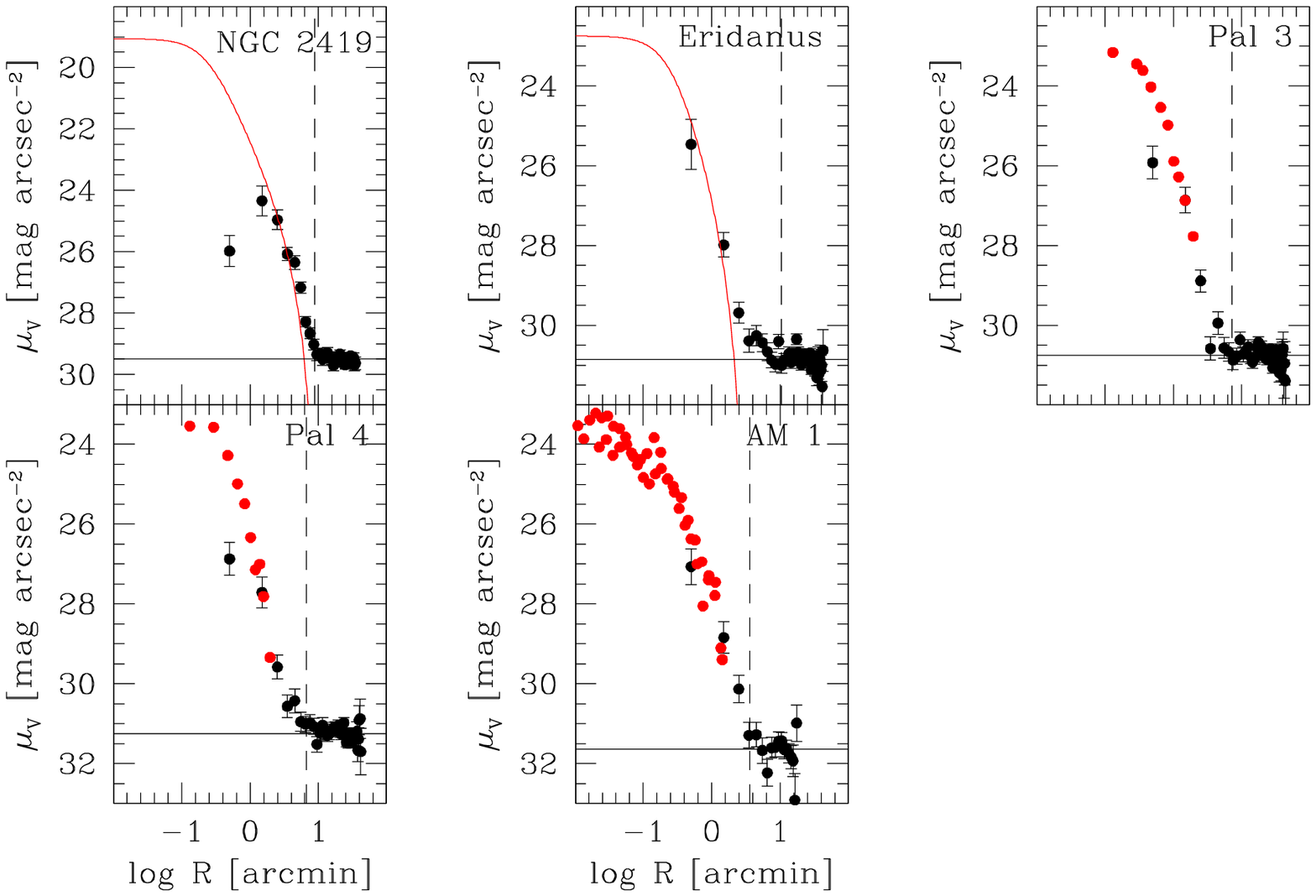}
 \caption{(Continued) Surface brightness profiles  of NGC 2419, Eridanus, Pal 3, Pal 4 and
 AM 1.}
\end{figure*}
  
\begin{figure}
 \includegraphics[width=8.6cm]{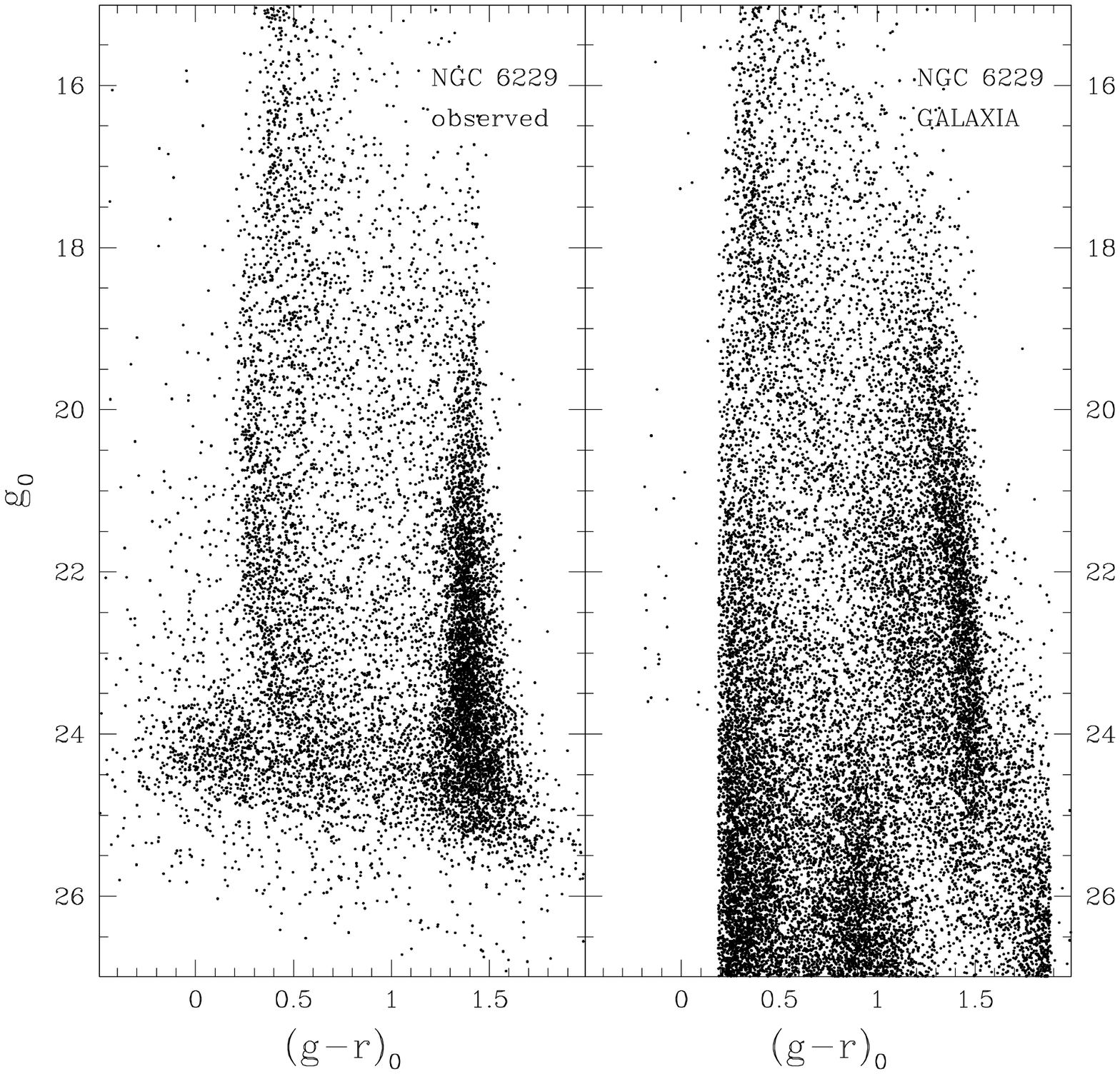}
 \caption{Comparison between the observed CMD of NGC 6229 (left panel) and 
 the synthetic CMD predicted by the \texttt{GALAXIA} model (right panel).}
\label{fig:cmdgal}
\end{figure}

\begin{figure*}
 \includegraphics[width=\textwidth]{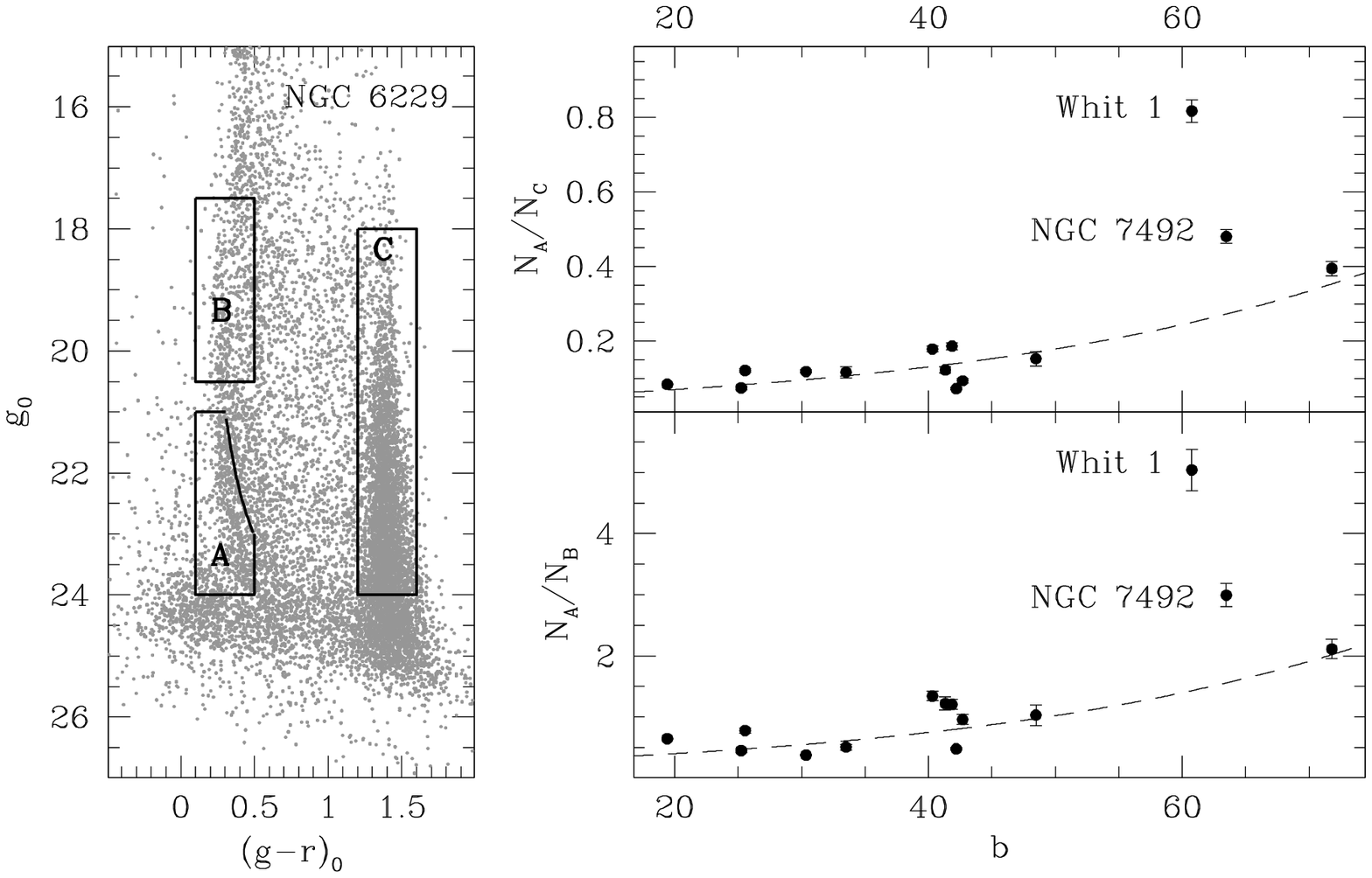}
 \caption{Left panel: CMD of NGC 6229. The adopted selection boxes are
 indicated. Right panels: number count ratios as a function of Galactic 
 latitude. The Galactic trend predicted by 
 the \texttt{GALAXIA} model is indicated in
 both panels with dashed lines.}
\label{fig:selb}
\end{figure*}
 
\begin{figure*}
 \includegraphics[width=\textwidth]{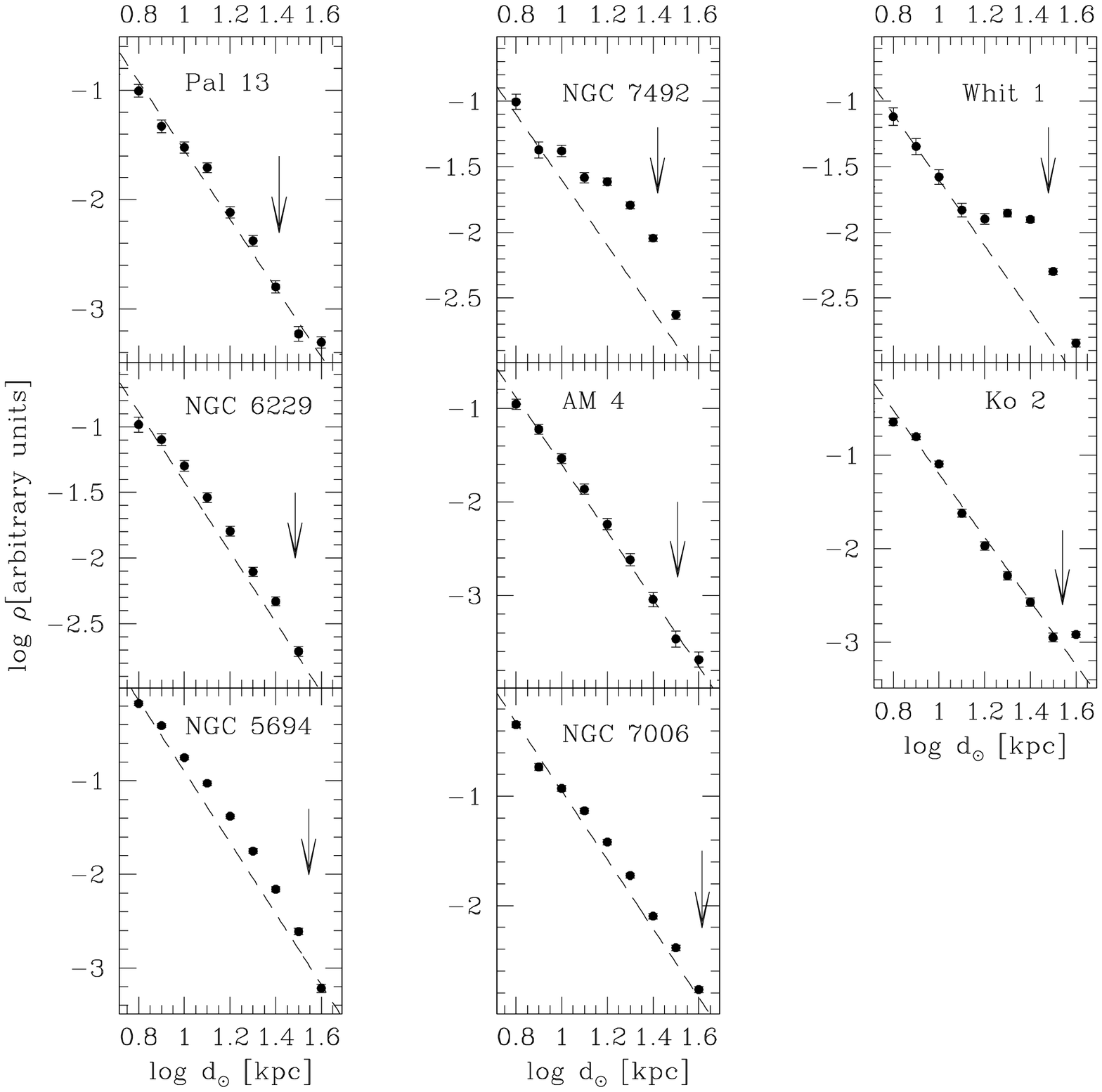}
 \caption{3D density profiles of the 8 GCs at heliocentric distances
 $d_{\odot}<45$ kpc. The power-law best fits are marked with dashed lines. The
 heliocentric distance of each cluster is indicated by arrows.}
\label{fig:dist}
\end{figure*}
 
\begin{figure*}
 \includegraphics[width=\textwidth]{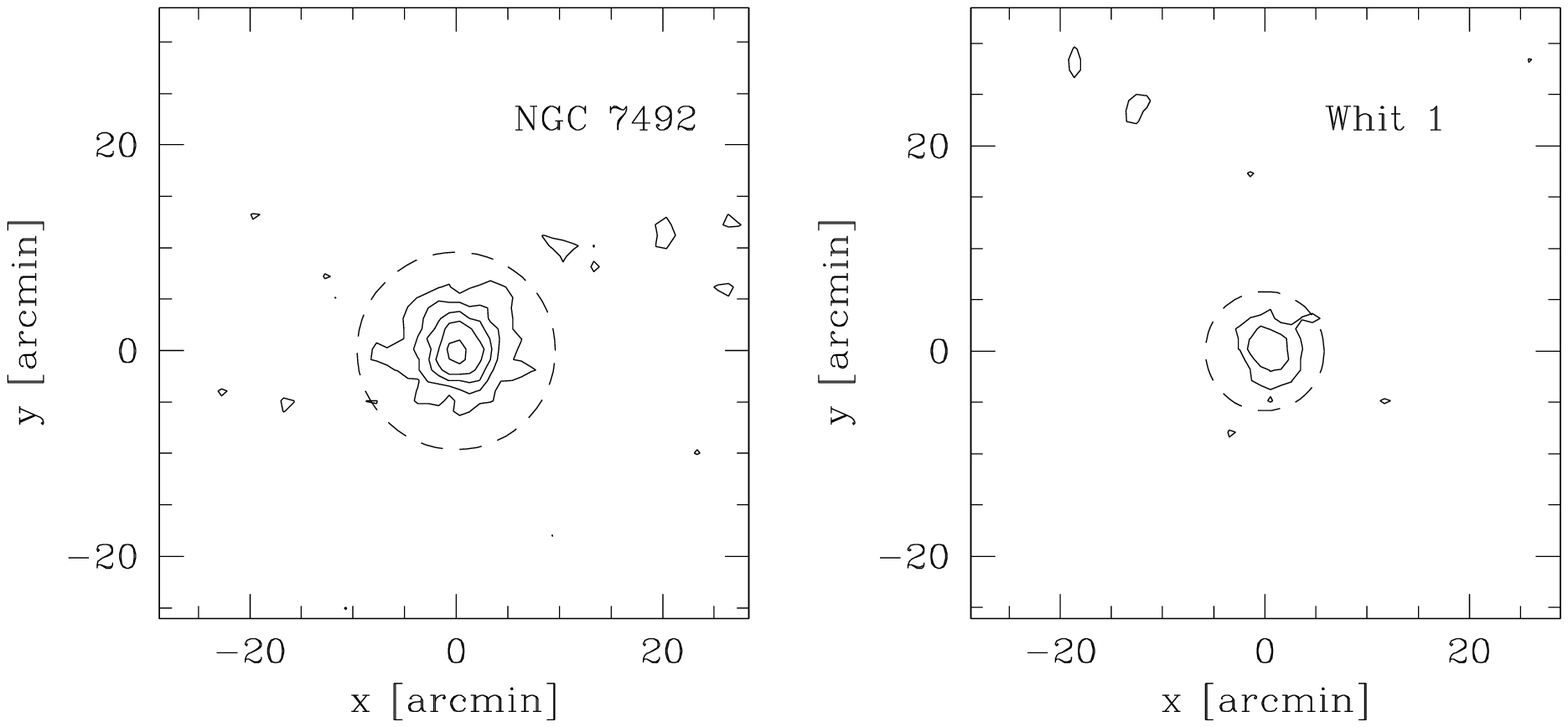}
 \caption{2D density maps of NGC 7492 (left panel) and Whiting 1 (right panel). 
 Density contours at 3$\sigma$ above the background are shown in logarithmic steps of 0.1.
 The adopted limiting radii $r_{lim}$ are marked with dashed lines.}
\label{fig:map2d}
\end{figure*}

The CMDs shown in Fig.~\ref{fig:cmd} were used to detect any
possible overdensity of stars compatible with that produced by the stellar
population of a surrounding remnant.

In the following analysis, magnitudes were dereddened  using the
reddening predicted by 
\cite{2011ApJ...737..103S}
 and the extinction
coefficients by \cite{1998ApJ...500..525S}.

To avoid the contamination from the stars of the GC itself, we only considered in our analysis  stars
located beyond a limiting radius $\rlim$, where the cluster contribution
vanishes.
For this purpose, we selected stars within 3 times the colour standard deviation about the MS
ridge line and counted the number of selected objects in concentric annuli.
The density profile was then derived by dividing the number counts by the area of each 
annulus. Number densities were then converted into magnitudes and
scaled to match the surface brightness profile of 
\cite{1995AJ....109..218T}.
 For those clusters not included in the Trager et al. (1995)
sample we adopted the best-fitting 
 \cite{1966AJ.....71...64K}  models provided by
\citet[Whiting 1 and AM 4]{2014MNRAS.445.2971C}, 
\citet[NGC 2419]{2011ApJ...738..186I} 
and \citet[Eridanus]{1996AJ....112.1487H}, 
which were normalised to match the
integrated $V$ magnitude listed in the \cite{1996AJ....112.1487H} catalogue. The resulting surface
brightness
profiles are shown in Fig.~\ref{fig:profile}. In all
clusters the profiles show the typical declining trend and flatten at a value
beyond which the Galactic contribution dominates over the clusters' stellar
population. For different choices of $\rlim$ we performed 
a $\chi^{2}$ test 
by selecting radial bins at projected distances
$R>\rlim$. We chose $\rlim$ as the minimum distance beyond which the
associated $\chi^{2}$
is compatible with a constant density at the 99.8\% confidence level. 
The mean surface brightness calculated at distances $R>\rlim$ was
then assumed
to be an upper limit for the surface brightness of the hypothetical underlying
remnant ($\muVmax$). Indeed, this approximate limit depends on
a number of assumptions. For example, a direct number count/surface brightness conversion 
was assumed, neglecting  possible differences in the stellar content of the
sampled population (\textsl{e.g.} age, metallicity, mass function, etc.). 
However, the cluster and its hypothetical host
galaxy are expected to be characterised by similar (old and metal-poor) stellar
populations at the same mean heliocentric distance (although with a 
larger spread for the remnant) and are 
affected by the same dust
extinction, thus sharing the same location in the CMD.
Note that while any stellar population with a surface
brightness brighter than $\muVmax$ would have produced a plateau 
at a surface brightness level larger than observed, the estimated background level could have been 
spuriously increased by ({\it i}) residual contamination from unresolved galaxies,
({\it ii}) Poisson fluctuations hampering the detection of any declining 
trend at low number counts, and ({\it iii}) the contribution of the remnants of 
other disrupted satellites lying along the line of sight to the GC. The adopted values of $\rlim$ and $\muVmax$
for the analysed GCs are listed in Table 2. The derived background
surface brightnesses lie in the range $29.4<\muVmax<31.7$ mag arcsec$^{-2}$ depending on the
sampled area and the cluster distance, and have typical uncertainties of $\sim$0.5 mag arcsec$^{-2}$.

Contamination from background galaxies has been significantly reduced by the strict cut in the 
sharpness parameter described in Sect. \ref{sec:obs}. While this criterion is
effective in removing extended objects at bright magnitudes, it cannot avoid
some
residual contamination at faint magnitudes ($g>24$) where the uncertainties in
the stellar profile determination produce an overlap between stars and galaxies
also in the sharpness parameter domain.
Moreover, because of such a strict criterion,
many faint stars at $g>24$ are removed causing a drop of the photometric
completeness. Because of this effect, a quantitative analysis was  
 only carried out for the subsample of 8 GCs 
within a heliocentric distance $d_{\odot}<45$ kpc. 

To quantitatively test for the presence of an overdensity of MS stars around the  
GCs of our sample we adopted three different methods:\\

First, we compared the number of stars contained in a region of the CMD
encompassing the predicted location of halo MS stars at the cluster distance 
with that predicted by the
\texttt{GALAXIA} Galactic model \citep[][method A]{2011ApJ...730....3S}. 
For this purpose, we counted the number of stars with colours within 3$\sigma$ 
about the cluster MS ridge line and within the magnitude interval
$20<g_{0}<24$ in both the observed ($N_\mathrm{obs}$) and the synthetic
($N_\mathrm{mod}$) CMD. In this range the photometric completeness is $>80\%$ and the
contamination from unresolved galaxies is negligible. The statistical significance of
any density difference was estimated as the number of standard deviations 
with respect to the zero difference hypothesis, assuming Poisson noise for 
number counts in the regime of large N ($\sigma(N)\sim\sqrt{N}$) and no 
covariance between $N_\mathrm{obs}$ and $N_\mathrm{mod}$
\begin{equation}
Q \; = \;
\frac{A_\mathrm{mod}N_\mathrm{obs}-A_\mathrm{obs}N_\mathrm{mod}}{\sqrt{A_\mathrm{mod}^{2}N_\mathrm{obs}+A_\mathrm{obs}^{2}N_\mathrm{mod}}}~~~~\mbox{\citet{2002A&A...381...51C}}
\end{equation}
where $A_\mathrm{obs}$ and $A_\mathrm{mod}$ are the sampled areas in the observed and
synthetic samples, respectively.
The values of $Q$ are listed in Table 2. 
A comparison between the observed CMD of NGC 6229 and the synthetic one predicted by the \texttt{GALAXIA} 
model in the same direction is shown in Fig. \ref{fig:cmdgal}, as an example. 
While on average number counts are
overestimated by the \texttt{GALAXIA} model, and yield negative
values for $Q$, highly significant overdensities ($Q>3$) are
 found around NGC~7492 and Whiting~1 (See Sec. \ref{sec:discussion}). 

These detections strongly depend on the adopted 
Galactic model. We hence also
consider two alternative model-independent approaches.
In the former approach (method B), we count the number of stars contained in three regions of the CMD defined as
follows (see Fig.~\ref{fig:selb}): 
\begin{itemize}
\item{A) A region defined by the
intersection between a selection box in the intervals $0.1<(g-r)_{0}<0.5$ and
$21<g_{0}<24$, and the portion of the CMD contained between the loci of a theoretical
isochrone of the \texttt{PARSEC-COLIBRI} database 
\citep{2017ApJ...835...77M}
 with $Z$=0.0005 and $t$=10 Gyr (hereafter referred to
``reference isochrone'') where the absolute magnitudes were 
 converted into the observational ones by adding the distance moduli
corresponding to distances of 20 and 45 kpc. This region
includes mainly halo MS stars in this interval of distances;}
\item{B) A region containing all stars in the intervals $0.1<(g-r)_{0}<0.5$ and
$17.5<g_{0}<20.5$. This region
includes mainly nearby thin and thick disc MS stars;}
\item{C) A region containing all stars in the intervals $1.2<(g-r)_{0}<1.6$ and
$18<g_{0}<24$. This region
includes low-mass MS stars of the Galactic disc(s).}
\end{itemize} 
Since these regions contain stars belonging to different Galactic
components, the number count ratios are expected to be functions of the Galactic
coordinates. In particular, in absence of any substructure, the larger the
Galactic latitude the smaller the portion of disc stars along the
line of sight. On the other hand, around clusters immersed in large 
overdensities, a significant increase above the mean trend is expected to be
noticeable. The number ratios $N_{A}/N_{B}$ and $N_{A}/N_{C}$ for the GCs of 
our sample as a
function of the Galactic latitude are shown in Fig.~\ref{fig:selb}. 
The majority of GCs define smoothly increasing relations when both number
count ratios are considered. Two clusters stray from these relations: NGC~7492
and Whiting~1, which both show an excess with respect to the global trend.

The third and last technique adopted to detect overdensities (method C) aims at studying the
3D density profile of halo MS stars along the line of sight of each GC. For this purpose, we
selected stars in the colour range $0.1<(g-r)_{0}<0.5$ and estimated their
corresponding absolute magnitude by interpolating along the reference isochrone.
The distance modulus (DM) of each star was then calculated and a histogram 
constructed $N(DM)$.
The corresponding 3D density profile becomes 
\begin{equation}
\log~\rho(d_{\odot}) \; = \; \log~N(DM) \, - \, 0.6 DM
\end{equation}
Here we are neglecting the effect of photometric errors and
of individual age/metallicity differences, which can spread the 
distribution up to 0.3 mag. For this reason, the above
density profile can only be used to study large-scale trends. However, the 
presence of a compact stellar population surrounding the cluster would produce
an increase of the local density at a distance similar to that of the cluster,
which should be noticeable in the density profile.
The derived 3D density profiles for the GCs of our sample are shown
in Fig.~\ref{fig:dist}. A power-law was fitted to the derived density 
profiles excluding the interval surrounding the cluster (at $\Delta log 
(d/\mathrm{kpc})<0.1$). Again, while in the majority of the analysed GCs the density
profile is well described by a single power-law, a bump at distances compatible
to those of the corresponding clusters is apparent 
around NGC~7492 and Whiting~1.
 
In Fig. \ref{fig:map2d} the two-dimensional density map of NGC 7492 and 
Whiting 1 are shown. These maps are constructed by applying the k-nearest 
neighbour density estimator \citep[][with k=10]{1986desd.book.....S} to the sample of stars 
with colours within 3$\sigma$ about the cluster MS ridge line. Note that, 
outside the adopted limiting radius, only barely significant overdensities 
homogeneously distributed across the field of view are apparent.

Summarising, the three independent methods outlined in this Section indicate
the presence of significant overdensities of MS stars around NGC~7492 and Whiting~1 
which are located in a small interval of distances compatible with those of these GCs. 
These stellar populations have a flat spatial distribution across the analysed field
of view 
extending far beyond the tidal radii (see Figs. \ref{fig:profile} and 
\ref{fig:map2d}) and present CMDs that are
inconsistent with those predicted by Galactic models. We therefore conclude
that the detected stellar populations are due to the remnants
of dwarf galaxies within which these two GCs are embedded. 
In particular, these two GCs are known to be located along the path of the 
Sagittarius stream, and the presence of the stellar population of this remnant 
in the surroundings of these two GCs has been revealed by previous works 
\citep{2014MNRAS.445.2971C,2017MNRAS.467L..91C,2018MNRAS.474.4766C}.

\section{Comparison with cosmological simulations}
\label{sec:models}
 
\begin{figure}
 \includegraphics[width=8.6cm]{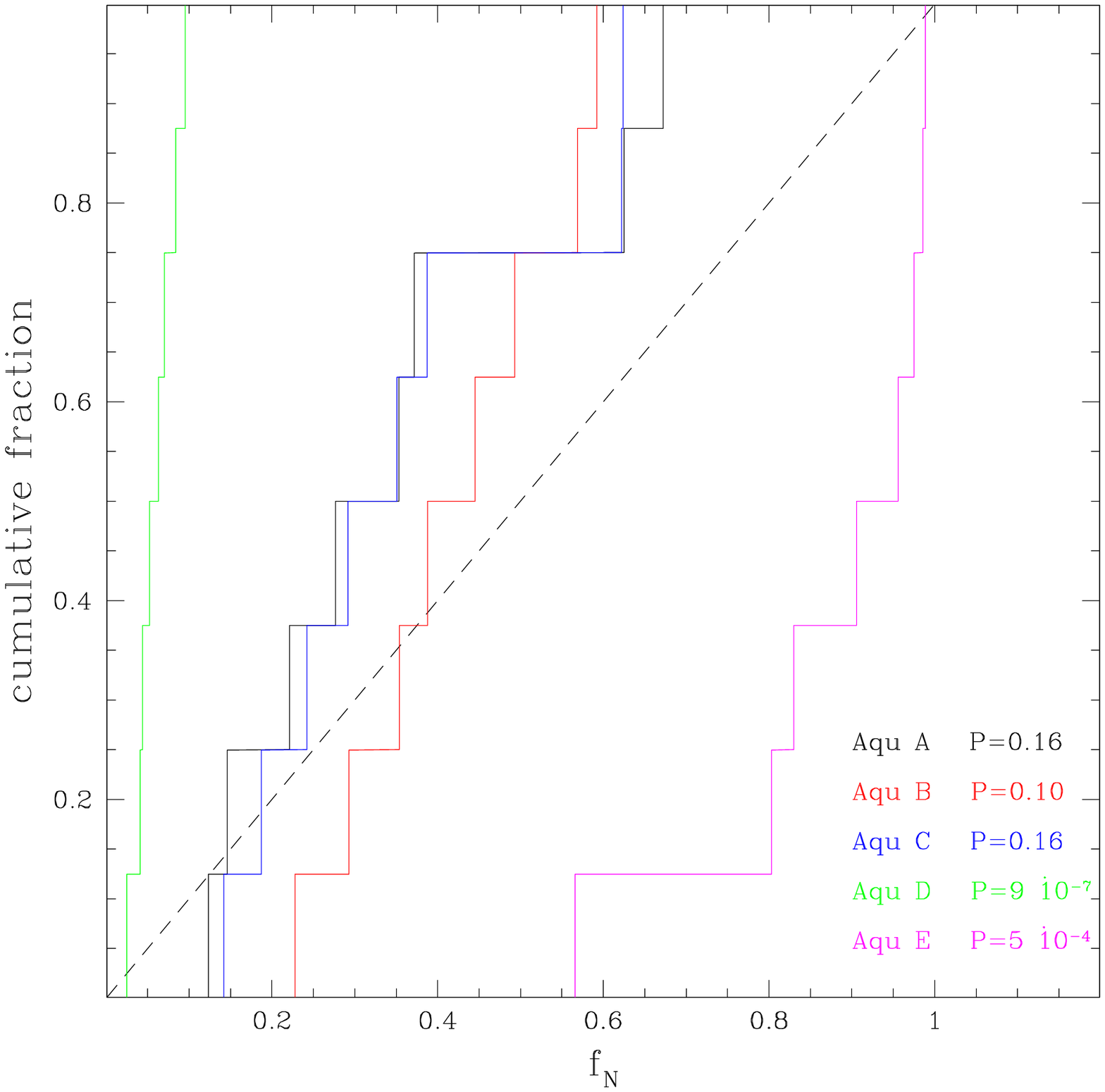}
 \caption{Cumulative distribution of percentiles in the number density domain
 for the 8 GCs at $d_{\odot} < 45$ kpc. Different lines refer to the various
 cosmological simulations. The probabilities associated with the KS test are also
 indicated.}
\label{fig:cumu}
\end{figure}
 
\begin{figure*}
 \includegraphics[width=\textwidth]{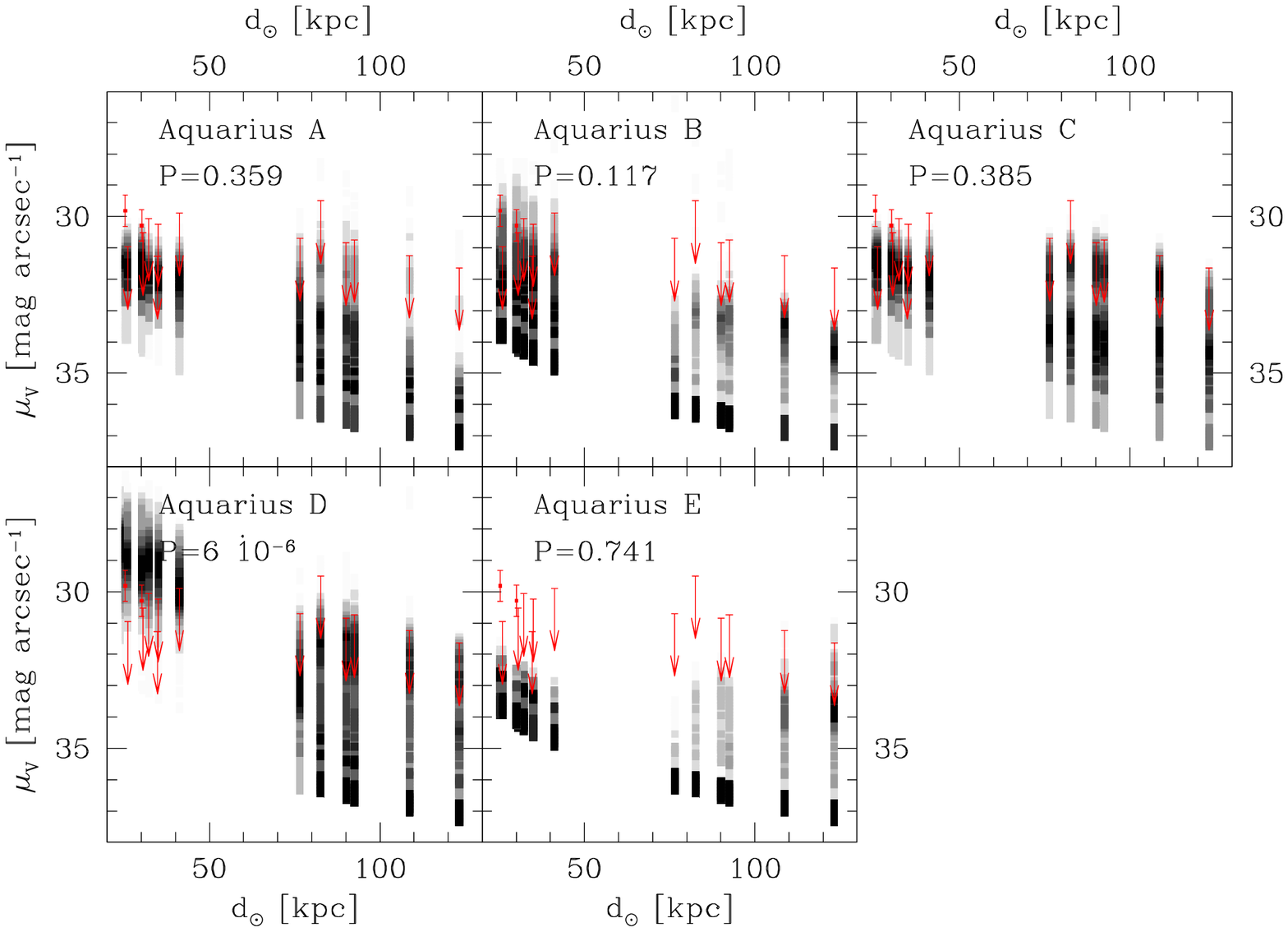}
 \caption{Distribution of extracted surface brightness as a function of the
 heliocentric distance. Darker regions indicate a higher number of extractions.
 The upper limits for the sample GCs are indicated by arrows.}
\label{fig:muv}
\end{figure*}
  
\begin{figure*}
 \includegraphics[width=\textwidth]{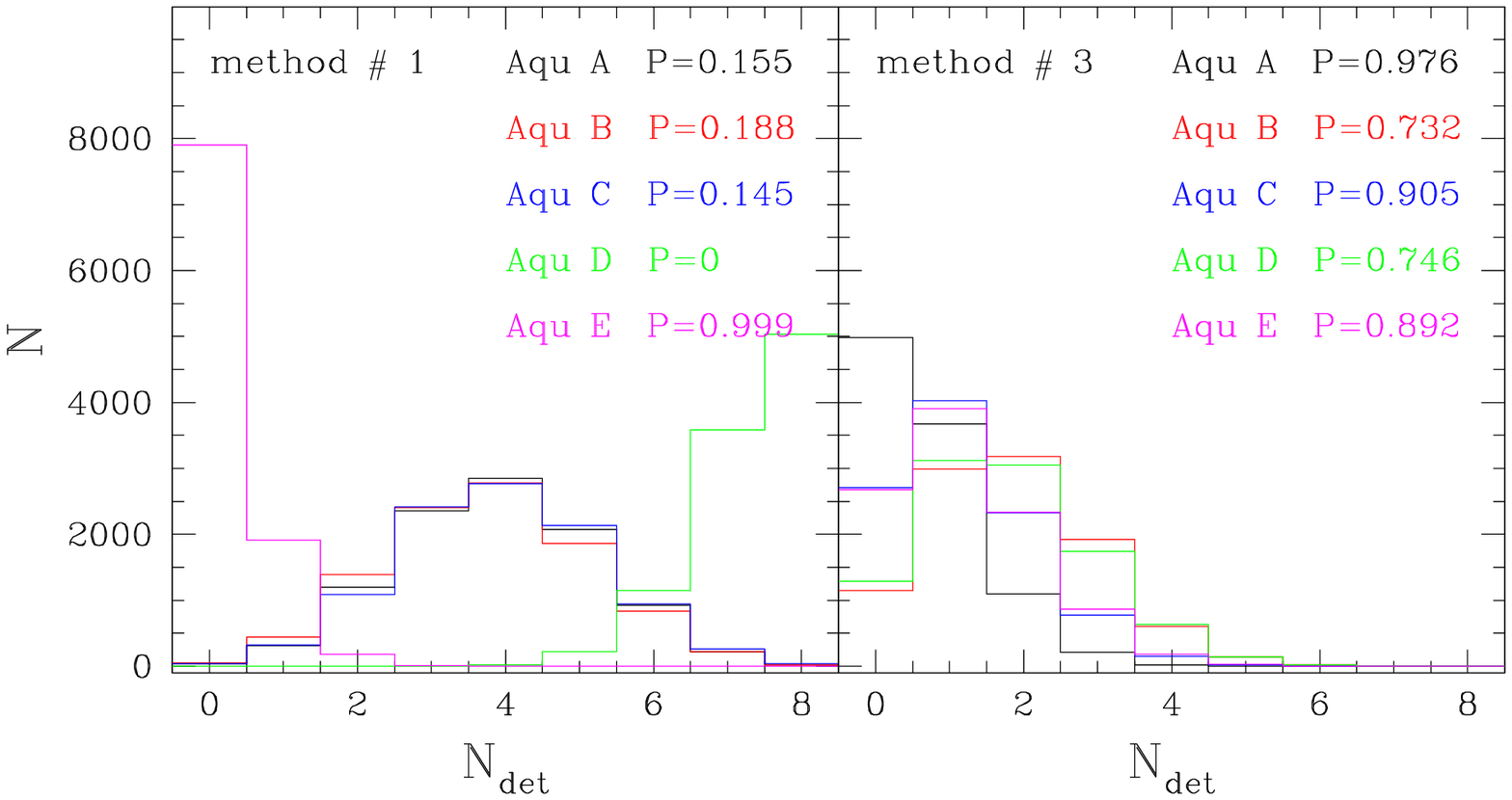}
 \caption{Histograms of detections in the sample of mock catalogues generated
 from the five Aquarius simulations. The left panel refers to method A while the
 right panel refers to method C. Different lines refer to the various
 cosmological simulations. The probabilities of detecting $\leq2$ underlying 
 remnants are also indicated.}
\label{fig:test}
\end{figure*}
  
\begin{figure*}
 \includegraphics[width=\textwidth]{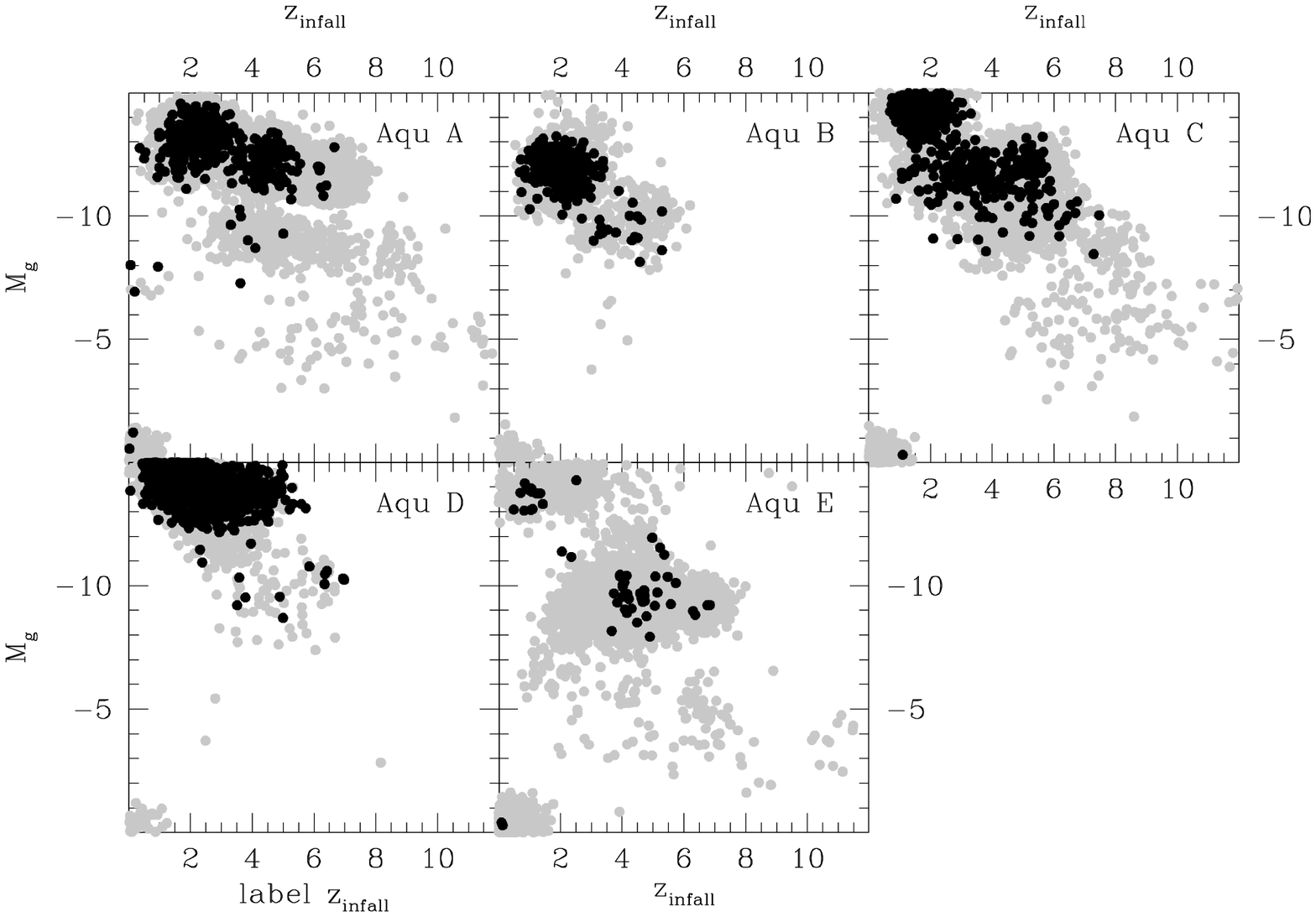}
 \caption{Distributions of detections in the sample of mock catalogues generated
 from the five Aquarius simulations in the infall redshift vs. host absolute $g$
 magnitude. Grey dots mark the location in this plane of all extractions,
 black dots mark the location of detections.}
\label{fig:zmg}
\end{figure*}

In the previous Section we presented evidence for significant overdensities
around 2 (out of 8) GCs at $d_{\odot}<45$ kpc, and provided upper limits for the surface
brightness of any undetected underlying stellar populations around each of the 14 GCs
of the entire sample. At first glance, this evidence could appear in 
contrast with the $\Lambda$CDM paradigm predicting that most of the Milky 
Way halo stars and GCs at large Galactocentric distances were accreted from
infalling satellites. Note, however, that the analysis performed in Sect.
\ref{sec:methods} is designed to detect overdensities above an average
background. Such a background could itself consist of by the superposition 
of the remnants of many satellites accreted during a Hubble time \citep{2015MNRAS.448L..77D}. So, the
existence of GCs with negative detections does not constitute a violation
of the $\Lambda$CDM model.
In this context, it is interesting to compare this result with the predictions 
of $\Lambda$CDM cosmological simulations.

We used the set of publicly available catalogues of synthetic stars by \cite{2015MNRAS.446.2274L}, 
which are based on the post-processing of the Aquarius project simulation \citep{2008MNRAS.391.1685S}.
These simulations assume the standard $\Lambda$CDM cosmology with parameters 
chosen to be consistent with the results from the WMAP 1-year data \citep{2003ApJS..148..175S}
and the 2dF Galaxy Redshift Survey data 
\citep{2001MNRAS.328.1039C}, 
 namely matter density parameter $\Omega_{M}$= 0.25; 
cosmological constant $\Omega_{\Lambda}$ = 0.75; power spectrum normalisation 
$\sigma_{8}$ = 0.9; spectral index $n_s$ = 1; and Hubble parameter $h$ = 0.73. 
Five Milky Way-like haloes (Aquarius A-E) were selected randomly from a set of isolated $\sim10^{12} M_{\odot}$ 
haloes identified in a lower resolution parent simulation of a 100 $h^{-1}$
Mpc$^{3}$ volume \citep{2008MNRAS.387..536G}.
The mass, age, and chemical composition of stars formed in each particle of the
simulation was calculated using the Durham semi-analytic model \texttt{GALFORM}, \citep{2000MNRAS.319..168C} 
which adopts physically motivated recipes to follow the chemical evolution of 
gas and stars including gas cooling and feedback effects. This model produces a satellite 
luminosity function that matches that of the Milky Way and also generates stellar 
populations in satellites that match the observed luminosity-metallicity 
relation.
A particle-tagging algorithm was then adopted to associate the stellar
component with six-dimensional phase-space volumes. As a final step, the total
number of stars associated with each tagged particle was calculated and associated
with a corresponding synthetic population (characterised by age, metallicity,
and magnitudes) using the technique described by
\cite{2012A&A...545A..14P} and the stellar isochrones of the \texttt{PARSEC}
 database \citep{2012MNRAS.427..127B}.

We tested the hypothesis that all the considered GCs were
originally part of accreted satellites. 
For each GC of our sample, we randomly extracted a particle from one of the above 
simulations among those formed $>10$ Gyr ago, belonging to the main halo at the
present day and at a Galactocentric distance
within $10\%$ of that of the considered GC, and marked it as a ``synthetic GC''.
With this choice we neglect all the complex and uncertain processes in the context 
of the formation and survival of GCs in different environments 
\citep[see \textsl{e.g.}][]{2010MNRAS.405..375G}, 
  implicitly assuming that the GC formation efficiency is 
constant in the satellites that formed the Galactic halo.
While this last assumption is clearly not suitable to study the general properties of GCs, 
it is not crucial for the purpose of quantifying the lumpiness of the halo.
All the particles of the simulation
were rotated in order to bring the synthetic GC at the same
Galactic coordinates of the considered cluster, adopting a
position for the Sun at
$(X,Y,Z)=(8,0,0)$ kpc. The distance moduli ($DM_{i}$) of the particles located within a projected
position on the sky of 1 square degree were 
calculated and added to their $g$ absolute magnitudes ($M_{g,syn}$), thus constructing a 
synthetic CMD 
$$g_\mathrm{syn}=M_{g,syn}+DM_{i}$$
The number $N_\mathrm{syn}$ of stars contained within 3$\sigma$ about the cluster MS ridge line and in
the magnitude range $3.92<g_\mathrm{syn}-DM_{GC}<6$ was counted, where $DM_{GC}$
is the distance modulus of the considered GC. To express the
projected density in units of surface brightness, the total $V$-band luminosity
of the stellar population was calculated as
\begin{equation}
M_{V,tot} \; = \; -2.5 \, \log\left(\sum_{i}10^{-0.4 M_{V,i}}\right) \; .
\end{equation}
The number $N_\mathrm{pop}$ of synthetic stars within 3$\sigma$ about the cluster MS ridge line and 
in the absolute magnitude range $3.92<M_{g}<6$ was similarly
computed. 
The corresponding surface brightness was then estimated as
\begin{equation}
\mu_{V,syn} \; = \; M_{V,tot} \, + \, DM_{GC} \, - \, 2.5 \log\left(\frac{N_{syn}}{N_{pop}}\right) \, + \, 17.78
\label{eq:mu}
\end{equation}
We do not include the observational effects (photometric errors, incompleteness 
and contamination from background galaxies). These effects are indeed hard to 
be properly quantified in our data and they are not expected to be significant 
in the magnitude range where our analysis is conducted ($g<24$).

This procedure was repeated $10^4$ times and the corresponding
distributions of $N_\mathrm{syn}$ and $\mu_{V,syn}$ were computed.
The number of stars contained within the same magnitude limits defined above was counted in the observed CMD and the density of stars per square degree
$\Sigma_\mathrm{obs}$ was
calculated by dividing by the sampled area. The fractions $f_{N}$ of random extractions
with $N_\mathrm{syn}<\Sigma_\mathrm{obs}$ and
the fractions $f_{\mu}$ with $\mu_{V,mod}>\muVmax$
were then computed. 

If the MS star density estimated around the sample of observed GCs 
were compatible with that predicted by models, the distribution of percentiles
would be uniform.
In Fig.~\ref{fig:cumu} the cumulative distributions of percentiles of the 8 GCs
at $d_{\odot}<45$ kpc with respect to the $N_\mathrm{syn}$ distribution of the five Aquarius simulations 
are shown. On average, the distribution of percentiles is 
consistent with a uniform distribution (although a vast range of lumpiness is covered by the 5 considered simulations), 
indicating a general agreement with model predictions for 3 out 5 simulations. 
A KS test indicates probabilities $0.1<P<0.2$ for these simulations. However,  
the simulations Aquarius D and Aquarius E show opposite trends with distributions of percentiles 
that are incompatible with observations.
Note that Aquarius D is the most massive halo among the five simulations of the
Aquarius sample, with a virial mass $M=2.52\times10^{12}~M_{\odot}$, exceeding the
current estimate for the Milky Way \citep{2017MNRAS.468.3428P}. It is
therefore likely that the number density of the halo is also overestimated in this
simulation. On the other hand, simulation Aquarius E is the one with the lowest fraction 
of stars in the diffuse halo at distances $\dgc>20$ kpc, 
with the majority of stars in this distance range contained in bound satellites.
For this reason, the number density outside satellites turns out to be lower than those in other haloes.

The same conclusion can be drawn when the entire sample of 14 GCs is considered in
the $\mu_{V}$ space. The distributions of extracted surface brightness from each
of the considered simulations are shown
as a function of heliocentric distance in Fig.~\ref{fig:muv}. We  note 
that the background surface brightness
distribution is peaked at values in the range $30<\mu_{V,syn}<35$ mag arcsec$^{-2}$ 
following a declining trend with distance. At large distances, the surface
brightness distribution becomes skewed with a tail towards bright 
values at $\mu_{V,syn}<34$ mag arcsec$^{-2}$. This is a consequence of the increasing degree of lumpiness of the
halo at large distances: GCs forming within satellites in an early stage of
dissolution are still surrounded by the compact remnant of their host galaxy.
The upper limits provided in Sect.~\ref{sec:methods} 
all lie  at 
surface brightness levels that are brighter than the synthetic distributions within 
 all simulations, the only exception being simulation Aquarius
D, which predicts a high surface brightness incompatible with the data of GCs at
$d_{\odot}<50$ kpc (see above). 
Since our estimates are only upper limits to the actual surface brightness levels, the
probability they are compatible with the distributions predicted by a given model
is 
\begin{equation}
P \; = \; \prod_{i=1}^{14} f_{\mu}
\end{equation}
The derived probabilities are in the range
$0.11<P<0.75$, compatible with no significant difference from the data, 
again with the 
exception of simulation Aquarius D, which has $P<10^{-5}$.

An alternative approach is to apply the same analysis performed on data to
the mock catalogues
generated from the Aquarius simulations. Note that, among the three different
techniques described in Sect.\ref{sec:methods}, only methods A and C
can be applied, while method B (based on the ratio of number counts in
regions populated mainly by halo or disk stars) cannot be used since the
Aquarius simulations do not include a disk population. 
The detection probability ($P_{i}$) of each of the 8 GCs at $d_{\odot}<45$ kpc
was estimated by applying these two techniques to each of the $10^{4}$ MonteCarlo 
synthetic CMDs constructed from each of the five Aquarius simulations. Then, a set of 8
random numbers ($\eta_{i}$) uniformly distributed between 0 and 1 was extracted and
associated with the "synthetic GCs". The number of detected "synthetic GCs" was 
calculated as the number of times the inequality $\eta_{i}<P_{i}$ is verified.
The above procedure was repeated $10^{4}$ times and the distribution of
detections was constructed (see Fig. \ref{fig:test}). It can be seen that while
method A tends to be more efficient than method C in detecting underlying 
remnants (at the
cost of being dependent on the adequacy of the GALAXIA model), in both
cases the probability of detecting $\leq2$ GCs is non negligible.

It is interesting to investigate the detection efficiency of our analysis as a function of
the characteristics of the progenitor host galaxies of GCs. To do
this, for each of the $10^{4}$ sets of "synthetic GCs" extracted from a given
Aquarius simulation (see
above) we selected those particles located within 1 sq. deg around the 
"synthetic GCs" and calculated the median redshift and absolute magnitude of
their former host galaxies at the moment of their infall within the Milky Way halo. 
In this case, we considered the entire sample of 14 GCs and assumed as
"detected" any remnant producing a surface brightness $\mu_{V,syn}<\mu_{V,lim}$
(see eq. \ref{eq:mu}).
In Fig.
\ref{fig:zmg} the distributions of detections in this diagram are shown for the
five Aquarius simulations. Note that detected remnants occupy only a portion of
this diagram at $z_{infall}<6$ and $M_{g}<-10$, with small variations among the
considered simulations. So, while it is not surprising that we were able to 
detect the
Sagittarius stream \citep[at $z_{infall}\sim1$,
$M_{g,0}\sim-13.8$][]{2010ApJ...714..229L}, no further bright satellite 
accretion seems to have contributed to the Galactic GC system in recent epochs in
this Galactocentric distance range.

\section{Discussion}
\label{sec:discussion}

By analysing the projected density of stars around 14 GCs populating the
outer halo of the Milky Way (at Galactocentric distances $d_{GC}>25$ kpc) we
find evidence of significant overdensities surrounding 
 two clusters, NGC
7492 and Whiting 1. No similar features were detected in any of the
other surveyed GCs.

The presence of such extended, compact and conspicuous stellar populations
around these two GCs is confirmed through three independent techniques and 
cannot be associated with either the clusters themselves
(because of their homogeneous spatial distribution across the field of view) or
with the Galactic field (because of their CMD morphology). They are therefore
compatible with the existence of a galactic remnant surrounding these GCs.
The presence of overdensities around these two GCs was 
suggested by \cite{2014MNRAS.445.2971C}
 who classified the detection around
Whiting 1 as ``probable'' and that around NGC 7492 as ``uncertain''. 

The $N$-body
simulation of \cite{2010MNRAS.408L..26P}  predicts the presence of the 
Sagittarius stream at a coincident distance in the same
portion of the sky occupied by Whiting 1 and NGC 7492, although only the former
has a compatible radial velocity. In a recent spectroscopic
follow-up, \cite{2017MNRAS.467L..91C,2018MNRAS.474.4766C} 
confirmed this hypothesis by
revealing the presence of peaks in
the velocity distribution of their sample that cannot be associated with the
Galactic field. From these studies, they identified the presence of the 
Sagittarius stream in the neighbourhood of these GCs although only Whiting~1 seems
to be kinematically linked to this substructure.

We do not confirm the detection of a conspicuous stream around NGC 2419 claimed 
by \cite{2003ApJ...596L.191N} \citep[see also][]{2011ApJ...731..119R} using star counts of blue horizontal branch stars and 
later identified as an RR Lyrae overdensity by \cite{2013ApJ...765..154D}.
Although at the distance of this remote GC it is difficult to sample the CMD at 
the turn-off level with a good level of completeness, the large number of blue 
horizontal branch stars and RR Lyrae would imply a surface brightness well 
above the limit reached by our data. On the other hand, at distances $d_{\odot}>100$ kpc, 
the turn-off region of any hypotetical stellar population would overlap with the 
locus of background galaxies, thus hampering any significant detection. 
So, whether the observed overdensity is linked to the Virgo stellar stream 
\citep[as hypothesised by][]{2009ApJ...701L..29C} or to the outer portion of 
the trailing arm of the Sagittarius stream \citep{2014MNRAS.437..116B}, they 
should lie beyond this distance.

The lack of detected streams around all the other GCs is, however, not surprising. 
Indeed, the comparison with 5 different Milky Way-like galaxies in the Aquarius 
cosmological simulation indicates that both the predicted mean stellar densities and 
the surface brightness of the halo at the distances of the surveyed GCs as
well as the predicted detection efficiencies of the techniques employed in our 
analysis are 
still compatible with those measured in our survey, in agreement with
what was found by \cite{2008ApJ...680..295B,
2009ApJ...698..567S,2011ApJ...738...79X}, and \cite{2016ApJ...816...80J}.
 The limiting surface brightness of our survey is estimated to be
$29.4<\muVmax<31.7$ mag arcsec$^{-2}$ depending on the cluster distance and spatial coverage,
and is about 2 mag brighter than those of typical satellites predicted by 
models in this range of distances. Moreover, a further complication arises at distances $>40$ kpc 
because of the noise due to the strong contamination from blue background
galaxies that overlap with the 
turn-off region of any subjacent stellar population.
Any possible discrepancy cannot be revealed by our
survey and its detection would require deeper data collected over larger fields of 
view (provided that an efficient star/galaxy discrimination is available at 
very faint magnitudes). 

Remarkably, in M31 the PAndAS survey revealed a wealth of compact
substructures at distances $d>30$ kpc from its centre, with almost all GCs
beyond 40 kpc being associated with some of these features  
\citep{2009Natur.461...66M,2010ApJ...717L..11M}.  A different picture would appear if an external view
of the Milky Way could be obtained in a survey of similar depth, with only a few GCs
aligned with the path of the Sagittarius stream. This indicates that, in spite of
their similar masses and morphologies, M31 experienced
a turbulent history characterised by many more accretion events than 
in the Milky Way.

The simulations used here for comparison include several
approximations that could make them less than
ideal in this context. For
instance, they do not include the contribution and the dynamical effect of
 the halo component formed
in-situ from the gas accreted in past mergers or the
baryonic disc(s).
Besides these intrinsic limitations of the simulations, the
analysis performed in Sect. \ref{sec:models} contains many approximations and
uncertain assumptions (number density-surface brightness conversion,
photometric completeness, stellar population synthesis, etc.) and neglects the
variation of the
GC formation/destruction efficiency in different host galaxies, thus making
a proper quantitative comparison between models and observations uncertain.

Further constraints based on wide-field surveys of the halo stellar content at 
faint magnitudes, like those provided by the future Large Synoptic Survey
Telescope \citep{2008SerAJ.176....1I}, are clearly needed for a sound estimate of the degree of lumpiness of the halo at
distances beyond 40 kpc.

\section*{Acknowledgments}
DMD and EKG gratefully acknowledge support by the Sonderforshungsbereich "The 
Milky Way System" (SFB 881, especially through subprojects A2 and A8) of the German 
Research Foundation (DFG).
R.R.M. acknowledges partial support from the BASAL PFB-$06$ project as well
as FONDECYT project N$^{\circ}1170364$.
JAC-B acknowledges financial support received from the CONICYT-Chile
FONDECYT Postdoctoral Fellowship 3160502, from the Ministry for the
Economy, Development, and Tourism's Programa Iniciativa Cient\'ifica
Milenio through grant IC120009, awarded to the Millennium Institute of
Astrophysics (MAS), and from CONICYT's PCI program through grant
DPI20140066. SGD acknowledges a
partial support from the NSF grants AST-1313422, AST-1413600, and AST-1518308.
This work is based on observations obtained with MegaCam, a joint project of 
CFHT and CEA/DAPNIA, at the CFHT observing programs 09AF03 (PI: Valls-Gabaud), 09AC07, 09BC02, 
and 10AC06 (PI: C{\^o}t{\'e}), on data products 
produced at TERAPIX and the Canadian Astronomy Data Centre, and on data 
gathered with the 6.5 meter Magellan Telescopes located at Las Campanas 
Observatory, Chile under the observing program 2010B-0472 (PI: Geha).
We warmly thank Michele Bellazzini for useful discussions.

\label{lastpage}

\end{document}